\documentclass[times,review,preprint,authoryear]{elsarticle}

\usepackage{jcomp}
\usepackage{framed,multirow}

\usepackage{amssymb}
\usepackage{latexsym}

\usepackage{url}
\usepackage{xcolor}
\definecolor{newcolor}{rgb}{.8,.349,.1}
\definecolor{darkgreen}{RGB}{0, 100, 0}

\usepackage{amsmath}
\usepackage{algorithm2e}
\usepackage{algorithmic}
\usepackage{subcaption}
\journal{Journal of Computational Physics}
\begin{document}

\verso{Sreevatsa Anantharamu and Krishnan Mahesh}

\begin{frontmatter}

\title{A parallel and streaming Dynamic Mode Decomposition algorithm with finite precision error analysis for large data}

\author[1]{Sreevatsa \snm{Anantharamu}}
\author[1]{Krishnan \snm{Mahesh}\corref{cor1}}
\cortext[cor1]{Corresponding author: 
  Tel.: +1-612-624-4175;  
  fax: +0-000-000-0000;}
\ead{kmahesh@umn.edu}

\address[1]{Department of Aerospace Engineering and Mechanics, Univeristy of Minnesota, Minneapolis, Minnesota 55455, USA}

\received{1 May 2013}
\finalform{10 May 2013}
\accepted{13 May 2013}
\availableonline{15 May 2013}
\communicated{S. Sarkar}

\begin{abstract}
 A novel technique based on the Full Orthogonalization Arnoldi (FOA) is proposed to perform Dynamic Mode Decomposition (DMD) for a sequence of snapshots. A modification to FOA is presented for situations where the matrix $A$ is unknown, but the set of vectors $\{A^{i-1}v_1\}_{i=1}^{N-1}$ are known. The modified FOA is the kernel for the proposed projected DMD algorithm termed, FOA based DMD. The proposed algorithm to compute DMD modes and eigenvalues i) does not require Singular Value Decomposition (SVD) for snapshot matrices $X$ with $\kappa_2(X) \ll 1/\epsilon_m$, where $\kappa_2(X)$ is the 2-norm condition number of the snapshot matrix and $\epsilon_m$ is the relative round-off error or machine epsilon, ii) has an optional rank truncation step motivated by round off error analysis for snapshot matrices $X$ with $\kappa_2(X) \approx 1/\epsilon_m$, iii) requires only one snapshot at a time, thus making it a 'streaming' method even with the optional rank truncation step, iv) consumes less memory and requires less floating point operations to obtain the projected matrix than existing projected DMD methods and v) lends itself to easy parallelism as the main computational kernel involves only vector additions, dot products and matrix vector products. The new technique is therefore well-suited for DMD of large datasets on parallel computing platforms. We show both theoretically and using numerical examples that for FOA based DMD without rank truncation, the finite precision error in the computed projection of the linear mapping is $O(\epsilon_m\kappa_2(X))$. The proposed method is also compared to existing projected DMD methods for computational cost, memory consumption and relative round off error. Error indicators are presented that are useful to decide when to stop acquiring new snapshots. The proposed method is applied to several examples of numerical simulations of fluid flow.
\end{abstract}


\end{frontmatter}



\section{Introduction}
Numerical simulations of both laminar and turbulent flows with $O(10^{5-10})$ degrees of freedom generate high-dimensional datasets comprising of velocity and pressure field at multiple time instants. These datasets can be low-dimensional when expressed in appropriate bases, often termed as `modes'. The high-dimensional dataset can be represented as a linear combination of few of these modes to reasonable accuracy. A review of different modal decomposition techniques to identify these modes for fluid flows is given in \cite{taira2017modal} and \cite{rowley2017model}.

Dynamic Mode Decomposition (DMD) is a data-driven modal decomposition technique that identifies a set of modes from multiple snapshots of the observable vectors (defined in section \ref{sec:projdmdsection}). Each of these modes are assigned an eigenvalue which denotes growth/decay rate and oscillation frequency of the mode. The obtained modes and corresponding eigenvalues together capture the dynamics of the underlying system. \cite{rowley2009spectral} related the DMD modes and DMD eigenvalues to the eigenfunctions and eigenvalues of the Koopman operator. The Koopman operator \citep{mezic2013analysis} is an infinite dimensional linear operator that describes the evolution of linear and nonlinear dynamical systems. These connections make DMD applicable to nonlinear systems such as those governed by the Navier--Stokes equations. 

DMD for sequential snapshots uniformly separated in time was first introduced by \cite{schmidaps2008}. \cite{schmid2010dynamic} proposed two algorithms which used different choices of basis vectors to perform Galerkin projection of the assumed linear mapping between the snapshots. One algorithm was based on Arnoldi method with no orthogonalization, with the snapshot vectors as the basis vectors, and the other relied on singular value decomposition (SVD) of the sequence of snapshots, and used the resulting left singular vectors as the basis vectors. The SVD based method was seen to have better finite precision accuracy than the Arnoldi based method. \cite{tu2014dynamic} modified the SVD version of \cite{schmid2010dynamic} proposed for a sequence of snapshots, to consider snapshot pairs. They also proposed Exact DMD where the DMD modes and eigenvalues are defined as the eigendecomposition of the minimum Frobenius norm mapping that relates the snapshot pairs in a least-squares sense. This essentially involved a new definition of the DMD modes from the eigenvalues and eigenvectors of the projected linear mapping. The drawback of the SVD based DMD method discussed in \cite{schmid2010dynamic} and \cite{tu2014dynamic} (both projected and exact DMD) is that it requires access to all snapshots at once to compute the projected linear mapping. \cite{hemati2014dynamic} proposed a streaming version of \cite{tu2014dynamic} to process large and streaming datasets which requires access to only the current snapshot pair to compute the projected linear mapping, and also a compression procedure to reduce the effect of noise in the dataset on the DMD modes. However, as shown in this paper through numerical experiments, the finite precision error in computing the projected linear mapping from the streaming version of \cite{hemati2014dynamic} without compression can be large. Two DMD algorithms that are the same in theory might can have very different finite precision error. It is therefore important to have an estimate of finite precision error of the computed projection while devising DMD algorithms. A parallel version of the DMD algorithm proposed in \cite{schmid2010dynamic} was presented in \cite{sayadi2016parallel}. Scaling of the algorithm was shown upto 1024 processors. However, this method is not streaming as it requires access to all snapshot vectors to compute the projected linear mapping. The underlying parallel algorithm used the TSQR algorithm of \cite{demmel2012communication} to compute QR factorization of the snapshot vectors followed by SVD of a small upper triangular matrix. Several other variants of DMD apart from projected DMD and Exact DMD \citep{tu2014dynamic} include optimized DMD \citep{chen2012variants}, extended DMD \citep{williams2015data}, kernel DMD \citep{williams2014kernel}, noise corrected DMD \citep{dawson2016characterizing}, forward-backward DMD \citep{dawson2016characterizing}, total least-squares DMD \citep{dawson2016characterizing}, recursive DMD \citep{noack2016recursive}, sparsity promoting DMD \citep{jovanovic2014sparsity} and optimal mode decomposition \citep{wynn2013optimal}. 

Numerical experiments on the error due to the number of snapshots and spatial resolution were performed by \cite{duke2012error} for synthetically generated data with noise, and that due to the choice of observables is presented in \cite{zhang2017evaluating}. Approximate solution to eigenvalue problems using Arnoldi's method and Galerkin projection (\cite{saad2011numerical}) also provide estimates for the error associated with each eigenvector and eigenvalue pair. We utilize these results which have been used for numerical solution of eigenvalue problems in the context of DMD where the linear mapping is not explicitly known.

The contribution of this paper is a novel parallel streaming DMD algorithm suitable for large data along with its finite precision error analysis. DMD is inherently not a backward stable (refer \cite{higham2002accuracy} for more on backward stability) procedure to compute projection of linear mapping using orthonormal basis vectors. This is because it relies on the snapshot vectors and not on the knowledge of linear mapping $A$. This leads to dependence of the finite precision error in the computed projection on the condition number of the snapshot matrix. Also, this warrants care in the design of DMD algorithms which might otherwise lead to dependence on higher powers of condition number and subsequently large finite precision error. In this paper, we examine the finite precision arithmetic properties of the proposed and few other exisiting DMD methods. The proposed method allows quantitative estimates of whether additional snapshots are needed, and the finite precision error contributions to the estimated modes. These properties are obtained in a readily parallelizable and streaming formulation, which makes reliable DMD representation of large datasets possible.  

The proposed method is derived from the Full Orthogonalization Arnoldi (FOA) procedure used to compute the projection of a given linear mapping $A$ onto the Krylov subspace. However, the FOA procedure explicitly requires the knowledge of operating $A$ onto the successively generated Arnoldi vectors $v_j$. We avoid this by reformulating FOA such that in the $j^{th}$ step, the new method relies only on the additional knowledge of $A^{j}\psi_1$ to compute $Av_j$, where $\psi_1$ is a random starting vector used in FOA. This results in a streaming DMD algorithm which requires access to only the current snapshot vector $\psi_{j+1}:=A^{j}\psi_1$. 

The paper is organized as follows. Section \ref{sec:projdmdsection}, discusses the proposed method in the broader context of projected DMD methods. Section \ref{sec:foabaseddmd} presents the proposed FOA based DMD. First, the algorithm is derived in batch processed form by drawing parallels to the Rayleigh-Ritz procedure using FOA in section \ref{sec:foabaseddmdalgobpform}. Then, the method is recast in streaming form. Properties of the algorithm which include computational cost, memory consumption and finite precision error in computation of the projected linear mapping are discussed and compared to exiting projected DMD algorithms in sections \ref{sec:foabaseddmdpropcandm} and \ref{sec:fpefoabaseddmd}. Error indicators for DMD eigenvalues and eigenvectors computed using FOA based DMD and their finite precision quality are discussed in section \ref{sec:foabaseddmderrorind}. Parallel scaling of the proposed algorithm is presented in section \ref{sec:parallelscaling}. Section \ref{sec:numericalexperiments} demonstrates FOA based DMD algorithm for three test cases: i) linearized channel flow at $Re=10000$, ii) flow over a circular cylinder at $Re=100$ and iii) jets in cross-flow at two different jet velocity to cross flow velocity ratios.

\section{Projected DMD methods: background} \label{sec:projdmdsection}

Let $x_j\in\mathbb{C}^n$ be the state of system at time $t_j$. The states $x_j$ are assumed to be equally spaced in time. i.e. $t_j=(j-1)\Delta t$ where $\Delta t$ is the time separation between successive states. Let $\psi(x):=[\psi^1(x)\quad\psi^2(x)\quad\dots\quad\psi^M(x)]^T$ be a vector of functions i.e. each $\psi^i(x):\mathbb{C}^n\rightarrow\mathbb{C}$. Each $\psi^i(x)$ is termed an observable \citep{rowley2009spectral}, and $\psi(x)$ is termed vector of observables. The snapshot vector $\psi_j\in\mathbb{C}^M:= \psi(x_j)$. i.e. $\psi_j=[\psi^1(x_j)\quad\psi^2(x_j)\quad\dots\quad\psi^M(x_j)]^T$. Let $X_i^j\in \mathbb{C}^{M\times (j-i+1)}$ be the snapshot matrix formed by stacking the snapshot vectors from time $t_i$ to $t_j$:
\begin{equation}
    X_i^j := [\psi_i\quad \psi_{i+1} \quad \dots \quad \psi_{j-1} \quad \psi_j].
\end{equation}
A common state vector is the velocity at all points in the domain of fluid simulation and a common vector of observables is the state vector itself. Let $N$ denote the total number of snapshots collected. A linear mapping $A$ is assumed to relate the successive snapshot vectors as
\begin{equation}
    \psi_{i+1} = A\psi_i,\quad X_2^{N} = AX_1^{N-1}.
\end{equation}
%

As discussed by \cite{schmid2010dynamic}, for linear time evolution of the observable vectors, no assumption is involved. For nonlinear time evolution of the observable vectors, for $N \le R+1$ where $R$ is the rank of the snapshot matrix, a linear mapping can always be defined to connect the snapshot vectors over the time interval $[0,N\Delta t]$. However, such a linear mapping may only be an aproximation to then nonlinear evolution of the observable vectors. In section \ref{sec:reconerrorandinterp}, we discuss that the projected linear mapping $A$ contains information of the interpolant through the sampled observable vectors which approximates the nonlinear time evolution of the system. A linear mapping between the successive snapshots implies that the range of matrix $X_1^{N-1}$ is also a Krylov subspace associated with the assumed linear mapping $A$ and the starting vector $\psi_1$,
\begin{equation} \label{eq:krylovsubspace}
    K_{N-1}(A,\psi_1) := span\{\psi_1,A\psi_1,...,A^{N-2}\psi_1\}.
\end{equation}
The columns of $X_1^{N-1}$ can be linearly dependent and need not always form the set of basis vectors for the above Krylov subspace. If they are linearly dependent, then $N$ is modified to include the linearly independent snapshot sequence only.

The Galerkin statement for the approximate eigenvalue problem is to find $\lambda \in \mathbb{C}, v\in K_{N-1}(A,\psi_1)$ pair s.t. $Av-\lambda v$ is orthogonal to all the vectors in the subspace $K_{N-1}(A,\psi_1)$ in the $\ell^2$ inner product. Suppose span of columns of $Q\in \mathbb{C}^{M\times N-1}$ and $V\in \mathbb{C}^{M\times N-1}$ are two sets of basis vectors for $K_{N-1}(A,\psi_1)$, then the approximate eigenvector $v$ can be written as $Qz$ for $z\in \mathbb{C}^{N-1}$. The Galerkin problem for $\lambda \in \mathbb{C}, z \in \mathbb{C}^{N-1}$ pair is then the generalized eigenvalue problem,
\begin{equation} \label{eqn:pdmdgalprob}
\begin{split}
V^H\left(AQz-\lambda Qz\right) = 0, \\
V^H AQz = \lambda V^HQz.
\end{split}
\end{equation}
If $V=Q$, then we obtain,
\begin{equation}
Q^HAQz = \lambda Q^HQz.
\end{equation}
Furthermore, if columns of $Q$ are orthonormal, we obtain
\begin{equation} \label{eqn:pdmdgalproborthbas}
Q^H AQz = \lambda z.
\end{equation}
The goal of all projected DMD methods is to compute $Q$ (which may or may not be orthonormal) and $Q^H A Q$ from the snapshot vectors $X_1^N$ alone without knowledge of the linear mapping $A$. Some possible choices of $Q$ are : matrix of snapshots $X_1^{N-1}$ (\cite{schmidaps2008,schmid2010dynamic,rowley2009spectral}), left singular vectors of economy SVD of $X_1^{N-1}$ (\cite{schmid2010dynamic,sayadi2016parallel}) and orthonormal matrix from QR factorization of $X_1^{N-1}$ (\cite{hemati2014dynamic}).
\section{FOA based DMD} \label{sec:foabaseddmd}


\subsection{Algorithm} \label{sec:foabaseddmdalgobpform}

\begin{algorithm}
\caption{FOA based DMD in batch processed form.}
\label{algo:foabaseddmd}
\begin{algorithmic}[1]
\STATE Collect $N$ snapshots and form $X_1^N$. Create space for matrices $V_1^{N} \in \mathbb{C}^{M\times N}$, $\bar{H}_N\in\mathbb{C}^{N,N-1}$ and $\beta_N\in\mathbb{C}^{N,N}$ and initialize to 0.
\STATE Construct initial vector $v_1$ from the first snapshot, $v_1:=\frac{\psi_1}{\|\psi_1\|_2}$
\STATE $\beta_{1,1}=\|\psi_1\|_2$
\FOR{j=1 to N-1}
\STATE $\beta_{1,j+1}=h_{1,1:j-1}\beta_{1:j-1,j}$
\FOR{i=2 to j}
\STATE $\beta_{i,j+1}=h_{i,i-1:j-1}\beta_{i-1:j-1,j}$
\ENDFOR
\STATE $w=\frac{1}{\beta_{j,j}}\left(\psi_{j+1}-\sum_{i=1}^j\beta_{i,j+1}v_i\right)$
\STATE $h_{1:j,j} = {V_1^j}^Hw$
\STATE $w=w-V_1^jh_{1:j,j}$
\STATE $s_{1:j} = {V_1^j}^Hw$
\STATE $h_{1:j,j}=h_{1:j,j}+s_{1:j}$
\STATE $w=w-V_1^js_{1:j}$
\STATE $h_{j+1,j}=\|w\|_2;v_{j+1}=\frac{w}{h_{j+1,j}}$
\FOR{i=1 to j+1}
\STATE $\beta_{i,j+1} = \beta_{i,j+1} + h_{i,j}\beta_{j,j}$
\ENDFOR
\ENDFOR
\STATE Define $H_{N-1}:=\bar{H}_N(1:N-1,1:N-1)$.
\IF {rank truncation}
 \STATE Compute SVD of $\beta_{N-1}:=\beta_N(1:N-1,1:N-1)$, i.e., $\beta_{N-1}=U\Sigma W^H$.
 \STATE Choose the truncated rank $r$ and $U_r:=U(:,1:r)$ and form $U_r^HH_{N-1}U_r$.
 \STATE Compute right eigenvectors $\{z_i\}_{i=1}^{r}$ and eigenvalues $\{\lambda_i\}_{i=1}^{r}$ of $U_r^HH_{N-1}U_r$.
 \STATE DMD modes are $\{V_1^{N-1}U_rz_i\}$ and DMD eigenvalues are $\{\lambda_i\}_{i=1}^{r}$.
\ELSE
 \STATE Compute right eigenvectors $\{z_i\}_{i=1}^{N-1}$ and eigenvalues $\{\lambda_i\}_{i=1}^{N-1}$ of $H_{N-1}$.
 \STATE DMD modes are $\{V_1^{N-1}z_i\}_{i=1}^{N-1}$ and DMD eigenvalues are $\{\lambda_i\}_{i=1}^{N-1}$.
\ENDIF
\end{algorithmic}
\end{algorithm}

The batch processed form (i.e. all snapshots processed at once) of the proposed FOA based DMD is shown in Algorithm \ref{algo:foabaseddmd}. $V_1^N$ is the matrix formed by stacking the set of vectors $\{v_i\}_{i=1}^N$ as columns, and $h_{i,j}$ and $\beta_{i,j}$ are the entries of the matrices $\bar{H}_N$ and $\beta_N$ respectively. The algorithm takes snapshot matrix $X_1^N$ as input and computes DMD modes and eigenvalues with or without rank truncation. Rank truncation before computing DMD modes reduces the finite precision error in the computed projection of the linear mapping $A$, and is useful for snapshot matrices with $\kappa_2(X_1^{N-1})\approx 1/\epsilon_m$, where $\kappa_2(X_1^{N-1})$ is the 2-norm condition number defined as $\|X_1^{N-1}\|_2\|X_1^{{N-1}^{\dagger}}\|_2$ to . For snapshot matrices with condition number $\kappa_2(X_1^{N-1}) \ll 1/\epsilon_m$, no rank truncation is required before computing the DMD modes and eigenvalues. We use Classical Gram Schmidt (CGS) with reorthogonalization as the orthogonalization kernel in FOA based DMD. This is because parallel implementation of CGS sends lesser number of messages than Modified Gram Schmidt (MGS), and CGS makes use of matrix vector products which are more efficient to compute. Reorthogonalization ensures orthgonality of $V_1^N$ upto machine precision \citep{giraud2005rounding}. 

\begin{algorithm}
\caption{Rayleigh Ritz procedure with Arnoldi's method.}
\label{algo:rayleighritzarnoldi}
\begin{algorithmic}[1]
\STATE Given $A$. Create space for matrices $V_1^{N} \in \mathbb{C}^{M\times N}$ and $\bar{H}_N\in\mathbb{C}^{N,N-1}$.
\STATE Construct initial vector $v_1$ from the first snapshot. $v_1:=\frac{\psi_1}{\|\psi_1\|_2}$
\FOR{j=1 to N-1}
\STATE $w=Av_j$
\STATE $h_{1:j,j} = {V_1^j}^Hw$
\STATE $w=w-V_1^jh_{1:j,j}$
\STATE $s_{1:j} = {V_1^j}^Hw$
\STATE $h_{1:j,j}=h_{1:j,j}+s_{1:j}$
\STATE $w=w-V_1^js_{1:j}$
\STATE $h_{j+1,j}=\|w\|_2;v_{j+1}=\frac{w}{h_{j+1,j}}$
\ENDFOR
\STATE Define $H_{N-1}:=\bar{H}_N(1:N-1,1:N-1)$.
\STATE Compute right eigenvectors $\{z_i\}_{i=1}^{N-1}$ and eigenvalues $\{\lambda_i\}_{i=1}^{N-1}$ of $H_{N-1}$.
\STATE Approximate eigenvectors are $\{V_1^{N-1}z_i\}_{i=1}^{N-1}$ and approximate eigenvalues are $\{\lambda_i\}_{i=1}^{N-1}$.
\end{algorithmic}
\end{algorithm}

The batch processed form of FOA based DMD can be derived from the Rayleigh-Ritz procedure using the Arnoldi method shown in Algorithm \ref{algo:rayleighritzarnoldi}. The Rayleigh-Ritz procedure computes approximate eigenvectors and eigenvalues of linear mapping $A$. However, we use a modified Arnoldi method where the matrix $A$ is unknown, but the set of vectors $\{\psi_i\}_{i=1}^{N}$ is known. Steps 2 to 10 of Algorithm \ref{algo:rayleighritzarnoldi} constitute the Arnoldi method which generates an orthonormal basis for the Krylov subspace, $K_{N-1}(A,\psi_1)$. The vectors $v_i$ generated from the Arnoldi method, in addition to being orthonormal satisfy the property \citep{saad2011numerical}
\begin{equation}\label{eq:avarnoldi}
\begin{split}
  AV_1^{N-1} &= V_1^N\bar{H}_N,\\
  AV_1^{N-1} &= V_1^{N-1}H_{N-1} + h_{N,N-1}v_Ne_{N-1}^H,
\end{split}
\end{equation}
where $H_{N-1}:=\bar{H}_N(1:N-1,1:N-1)$ is the projection of $A$ onto the Krylov subspace $K_{N-1}\left(A,\psi_1\right)$. i.e.,
\begin{equation}\label{eq:projaontokrylov}
{V_1^{N-1}}^HAV_1^{N-1}=H_{N-1},
\end{equation}
and hence the eigenvalues and eigenvectors of $H_{N-1}$ can be used to approximate those of $A$ (steps 13 and 14 of Algorithm \ref{algo:rayleighritzarnoldi}).

The Rayleigh Ritz procedure with the Arnoldi method cannot be directly used to perform projected DMD, as the step 4 in Algorithm \ref{algo:rayleighritzarnoldi} requires knowledge of the assumed linear mapping $A$, which is not explictly known before hand in DMD. However, we can compute $Av_j$ by using a linear combination of the snapshot vector $\psi_{j+1}$ and the set of vectors $\{v_i\}_{i=1}^j$. The Arnoldi vectors $v_i$'s generated by the above algorithm have the following property \citep{saad2011numerical},
\begin{equation}
span\{v_1,\dots,v_j\} = span\{\psi_1,\dots,\psi_j\} = span\{\psi_1,A\psi_1,\dots,A^{j-1}\psi_1\}.
\end{equation}
So, $\psi_j$ may be expressed as a linear combination of $\{v_i\}_{i=1}^j$,
\begin{equation}
\psi_j=\sum_{i=1}^j\beta_{i,j}v_i.
\end{equation}
In matrix form, 
\begin{equation}
\psi_j=V_1^j\beta_{1:j,j},
\end{equation}
where $V_1^j$ is the matrix formed by vectors $\{v_i\}_{i=1}^j$ as columns. Suppose, we are in the $j^{th}$ iteration of the Arnoldi method. We can develop a method to compute $Av_j$ from knowledge of $\bar{H}_{j}$, the non-zero entries of $j^{th}$ column of $\beta_N$ and the Arnoldi relation until the $(j-1)^{th}$ iteration using 
\begin{equation}
\begin{split} \label{eqn:avpsijp1beta}
\psi_{j+1}=A\psi_j&=AV_1^{j}\beta_{1:j,j} \\
&=AV_1^{j-1}\beta_{1:j-1,j} + Av_j\beta_{j,j}\\
&=V_1^j\bar{H}_j\beta_{1:j-1,j} + Av_j\beta_{j,j},
\end{split}
\end{equation}
where $\bar{H}_j$ is the $j\times (j-1)$ top left submatrix of $\bar{H}_N$. Rearranging,
\begin{equation}
Av_j=\frac{1}{\beta_{j,j}}\left(\psi_{j+1}-v_1\sum_{i=1}^{j-1}h_{1,i}\beta_{i,j}-\sum_{k=2}^jv_k\sum_{i=k-1}^{j-1}h_{k,i}\beta_{i,j}\right).
\end{equation}

Once the $j^{th}$ step of Arnoldi iteration is performed and the non-zero entries of the $j^{th}$ column of $\bar{H}_{N}$ are computed, the $\left(j+1\right)^{th}$ column of $\beta_N$ can be computed using the relation $Av_j=\sum_{i=1}^{j+1}h_{i,j}v_i$ in equation \ref{eqn:avpsijp1beta} which leads to
\begin{equation}
\begin{split}
\psi_{j+1}&=V_1^j\bar{H}_j\beta_{1:j-1,j} + V_1^{j+1}h_{1:j+1,j}\beta_{j,j}, \\
&=V_1^{j+1}\bar{H}_{j+1}\beta_{1:j,j}.
\end{split}
\end{equation}

The above equation shows that the non-zero entries in the $(j+1)^{th}$ column of $\beta_N$ can be computed as,
\begin{equation} \label{eqn:betacolumnupdate}
\begin{split}
\beta_{1,j+1}&=\sum_{i=1}^{j}h_{1,i}\beta_{i,j}, \\
\beta_{k,j+1}&=\sum_{i=k-1}^{j}h_{k,i}\beta_{i,j}\quad ;\quad k=2,\dots,j+1.
\end{split}
\end{equation}
The steps 1-19 of Algorithm \ref{algo:foabaseddmd} incorporating these modifications, constitutes what we call the modified FOA procedure.

Essentially, we are factorizing $X_1^N$ into $V_1^N$ which is a matrix with orthonormal columns obtained from the Arnoldi method, and $\beta_{N}$ which is an upper triangular matrix. Incorporating these changes yields the FOA based DMD shown in Algorithm \ref{algo:foabaseddmd} except the rank truncation part whose rationale will be discussed in section \ref{sec:fpefoabaseddmd}. Even though the Rayleigh Ritz procedure and FOA based DMD are theoretically equivalent, their computer implementations will yield different computed projection of $A$ whose error will be analyzed in \ref{sec:fpefoabaseddmd}.

The batch processed version of FOA based DMD in Algorithm \ref{algo:foabaseddmd} can be equivalently recast in streaming form shown in Algorithm \ref{algo:foabaseddmdstr}. The streaming update routine shown in Algorithm \ref{algo:foabaseddmdstrupdate} is called in each pass of the streaming algorithm where a single, or a set of snapshot vectors, are processed to a compute better projection of the assumed linear mapping between the successive snapshots. Specifically, in each pass of the `while loop' in Algorithm \ref{algo:foabaseddmdstr}, the snapshot matrix $X$ is formed by stacking $p$ new consecutive snapshot vectors as columns. Then, the streaming FOA based DMD update routine shown in Algorithm \ref{algo:foabaseddmdstrupdate} is called to update the matrix $H$, matrix $V$ with Arnoldi-generated orthonormal basis, the upper triangular matrix $\beta$, and $q$ which stores the current number of columns of $V$. Once these quantities are updated, the DMD modes and eigenvalues can be optionally evaluated with or without rank truncation. Usually, in streaming algorithms, $p$ is set to 1 and only one new snapshot vector is used to update the matrices. However, in Algorithm \ref{algo:foabaseddmdstrupdate}, snapshots can also be streamed in batches of $p$. This would reduce the number of times the matrices are resized (step 2 of Algorithm \ref{algo:foabaseddmdstrupdate}) while processing a fixed total number of snapshots. For a given total number of snapshots, the computed Arnoldi vectors, the projection of the linear mapping, and the upper triangular matrix computed using the streaming form, are exactly the same as those computed using the batch processed form since the exact same floating point operations are carried out in the same order, in both algorithms. So, unless otherwise stated, we refer to the batch processed form of FOA based DMD in the following sections. Also, from Algorithm \ref{algo:foabaseddmdstr}, note that FOA based DMD retains its streaming property even with rank truncation. This is because FOA based DMD with rank truncation additionally requires computing the SVD of $\beta_{N-1}$ and the new projected matrix $U_r^HH_{q-1}U_r$ which does not require previous snapshots. 
\begin{algorithm}
\caption{FOA based DMD in streaming form.}
\label{algo:foabaseddmdstr}
\begin{algorithmic}[1]
\STATE q=0.
\WHILE {true}
\STATE Collect $p$ new snapshot vectors in $X$.
\STATE $(H,V,\beta,q)$ = streamingFOAbasedDMDupdate $(X,p,H,V,\beta,q)$.
\IF {compute DMD modes and eigenvalues}
\STATE Define $H_{q-1}:=H_{1:q-1,1:q-1}$
\IF {rank truncation}
 \STATE Compute SVD of $\beta_{q-1}:=\beta_{1:q-1,1:q-1}$, i.e., $\beta_{q-1}=U\Sigma W^H$.
 \STATE Choose the truncated rank $r$ and $U_r:=U_{:,1:r}$.
 \STATE Compute right eigenvectors $\{z_i\}_{i=1}^{r}$ and eigenvalues $\{\lambda_i\}_{i=1}^{r}$ of $U_r^HH_{q-1}U_r$.
 \STATE DMD modes are $\{V_{:,1:q-1}U_rz_i\}_{i=1}^r$ and DMD eigenvalues are $\{\lambda_i\}_{i=1}^{r}$.
\ELSE
 \STATE Compute right eigenvectors $\{z_i\}_{i=1}^{q-1}$ and eigenvalues $\{\lambda_i\}_{i=1}^{q-1}$ of $H_{q-1}$.
 \STATE DMD modes are $\{V_{:,1:q-1}z_i\}_{i=1}^{q-1}$ and DMD eigenvalues are $\{\lambda_i\}_{i=1}^{q-1}$.
\ENDIF
\ENDIF
\IF {stopping criterion == true}
\STATE exit do while loop.
\ENDIF
\ENDWHILE
\end{algorithmic}
\end{algorithm}
\begin{algorithm}
\caption{Streaming FOA based DMD update routine.}
\label{algo:foabaseddmdstrupdate}
\begin{algorithmic}[1]
\STATE $(H,V,\beta,q)$ = streamingFOAbasedDMDupdate $(X,p,H,V,\beta,q)$
\STATE Expand $H$ to $\left(p+q\right)\times \left(p+q-1\right)$ Upper Hessenberg matrix, \\
$V$ to $M\times \left(p+q\right)$ matrix, \\
$\beta$ to $\left(p+q\right)\times \left(p+q\right)$ upper triangular matrix and set all new entries to 0.
\STATE $jb:=q$
\IF{$q==0$}
\STATE  $\beta_{1,1}=\|X_{:,1}\|_2$
\STATE $V_{:,1}:=\frac{X_{:,1}}{\|X_{:,1}\|_2}$
\STATE $jb:=1$
\ENDIF
\FOR{$j=jb$ to $p+q-1$}
\STATE $\beta_{1,j+1}=H_{1,1:j-1}\beta_{1:j-1,j}$
\FOR{$i=2$ to $j$}
\STATE $\beta_{i,j+1}=H_{i,i-1:j-1}\beta_{i-1:j-1,j}$
\ENDFOR
\STATE $w=\frac{1}{\beta_{j,j}}\left( X_{:,j-q+1} - \sum_{i=1}^j\beta_{i,j+1}V_{:,i}\right)$
\STATE $H_{1:j,j}={V_{:,1:j}}^Hw$
\STATE $w=w-V_{:,1:j}H_{1:j,j}$
\STATE $s_{1:j}={V_{:,1:j}}^Hw$
\STATE $H_{1:j,j}=H_{1:j,j}+s_{1:j}$
\STATE $w=w-V_{:,1:j}s_{1:j}$
\STATE $H_{j+1,j}=\|w\|_2;V_{:,j+1}=\frac{w}{H_{j+1,j}}$
\FOR{$i=1$ to $j+1$}
\STATE $\beta_{i,j+1} = \beta_{i,j+1} + H_{i,j}\beta_{j,j}$
\ENDFOR
\ENDFOR
\STATE $q=q+p$
\end{algorithmic}
\end{algorithm}

\subsection{Properties} \label{sec:foabaseddmdprop}

\subsubsection{Computational cost and memory consumption} \label{sec:foabaseddmdpropcandm}

This section presents the cost (number of floating point operations) and memory requirements of FOA based DMD, and compares it to existing projected DMD methods. We show in \ref{app:foabaseddmdcandm} that the cost of computing the projection of linear mapping $A$ onto the Krylov subspace $K_{N-1}(A,\psi_1)$ using snapshot matrix $X_1^{N}$ without rank truncation is approximately $5MN^2+N^3/3$, and with rank truncation is $5MN^2+37N^3/3+2rN^2+2r^2N$, where $r$ is the truncated rank. The approximate number of floating point operations for SVD based DMD (with and without rank truncation) of \cite{schmid2010dynamic} and streaming DMD of \cite{hemati2014dynamic} to obtain corresponding projected matrices is derived in \ref{app:svdbaseddmdcandm} and \ref{app:streamingdmdcandm}  respectively, and compared in Table \ref{tab:costcompnort}. Note that only terms which are cubic in the dimensions $M,N$ of the problem are retained. From Table \ref{tab:costcompnort}, it can be seen that FOA based DMD without rank truncation requires the least number of floating point operation count when compared to the existing projected DMD methods. We also compare the cost of rank truncated FOA based DMD and SVD based DMD in Table \ref{tab:costcompwrt}. Note that we do not consider the compressed streaming DMD of \cite{hemati2014dynamic} since this algorithm does not give the same result as rank truncated FOA based DMD and SVD based DMD methods (\ref{app:streamingdmdcandm}). Table \ref{tab:costcompwrt} shows that even with rank truncation, the cost of FOA based DMD is smaller than SVD based DMD. 

We also derive the memory (number of floating point numbers) required for the three projected DMD methods considered (both with and without rank truncation) in \ref{app:candm}. Table \ref{tab:memcompnort} shows that without rank truncation, the FOA based DMD method utilizes approximately half the space as the other two projected DMD methods. Also, when rank truncation is used, from table \ref{tab:memcompwrt} we see that FOA based DMD still uses less memory than the SVD based DMD method.
\begin{table}
\begin{center}
\begin{tabular}{ |c|c| } 
 \hline
 Method & Cost without rank truncation \\ 
 \hline
 FOA based DMD  & $5MN^2+N^3/3$ \\
 \hline
 SVD based DMD & $8MN^2+22N^3$ \\
 \hline
 Streaming DMD & $10MN^2+16N^3/3$ \\
 \hline
\end{tabular}
\end{center}
\caption{Cost (approximate) comparison of projected DMD methods without rank truncation.}
\label{tab:costcompnort}
\end{table}

\begin{table}
\begin{center}
\begin{tabular}{ |c|c| } 
 \hline
 Method & Cost with rank truncation \\ 
 \hline
 FOA based DMD  & $5MN^2+37N^3/3+2rN^2+2r^2N$ \\
 \hline
 SVD based DMD & $6MN^2+20N^3+2Mr^2+2r^3$ \\
 \hline
\end{tabular}
\end{center}
\caption{Cost (approximate) comparison of projected DMD methods with rank truncation.}
\label{tab:costcompwrt}
\end{table}

\begin{table}
\begin{center}
\begin{tabular}{ |c|c| } 
 \hline
 Method & Memory required without rank truncation \\ 
 \hline
 FOA based DMD  & $MN+N^2$ \\
 \hline
 SVD based DMD & $2MN+2N^2$ \\
 \hline
 Streaming DMD & $2MN+4N^2$ \\
 \hline
\end{tabular}
\end{center}
\caption{Memory requirement (approximate) comparison of projected DMD methods without rank truncation.}
\label{tab:memcompnort}
\end{table}

\begin{table}
\begin{center}
\begin{tabular}{ |c|c| } 
 \hline
 Method & Memory requied with rank truncation \\ 
 \hline
 FOA based DMD  & $MN+2N^2+r^2$ \\
 \hline
 SVD based DMD & $2MN+N^2+r^2$ \\
 \hline
\end{tabular}
\end{center}
\caption{Memory requirement (approximate) comparison of projected DMD methods with rank truncation.}
\label{tab:memcompwrt}
\end{table}


\subsubsection{Finite precision error in computed projected linear mapping} \label{sec:fpefoabaseddmd}

DMD computes projection of the assumed linear mapping $A$ generally through orthonormal basis vectors (as this leads to well-conditioned eigenvalue problem) which are generated using the snapshot vectors $\{A^{i-1}\psi_1\}_{i=1}^{N}$. Since, we do not know explicitly the mapping $A$, as we will see in this section, the finite precision error in the computed projection depends on the condition number of the snapshot matrix. However, depending on the algorithm used to compute the projection, the error in projection might even depend on different powers of the condition number and might lead to larger error. This makes finite precision error analysis important for DMD algorithms. However, such an issue is not present if we explicitly know the linear mapping $A$ and use the standard Arnoldi procedure in Algorithm \ref{algo:rayleighritzarnoldi}, as this leads to computed projection of $A$ which in a relative sense depends only on machine precision \citep{trefethen1997numerical}.

We first review the numerical stability results of QR factorization using CGS with reorthogonalization. Then, we perform finite precision error analysis of FOA based DMD and motivate rank trucation as a method to obtain small error due to finite precision arithmetic. Note that in addition to reducing the finite precision error, rank truncation also helps in rejecting statistical noise in the accumulated data. We also perform finite precision error analysis of SVD based DMD and streaming DMD and compare to FOA based DMD. Quantities with $\text{ $\hat{}$ }$ over them indicate computed counterparts in finite precision arithmetic and $\epsilon_m$ denotes machine precision. In deriving error estimates, we only consider real matrices and vectors. $c_i$'s are defined as constants that are moderate powers of $M,N$. i.e. $c_i:=c_i(M,N)$.

\paragraph{FOA based DMD}

The finite precision error analysis of QR factorization using CGS and MGS was performed by \cite{bjorck1992loss} and discussed in \cite{higham2002accuracy}. \cite{giraud2005rounding} showed that CGS and also MGS with reorthogonalization not only retain the backward stability of QR factorization, but also yield vectors which are orthonormal upto machine precision. The finite precision results of QR factorization using CGS with reorthogonalization to orthonormalize the vectors obtained by \cite{giraud2005rounding} is mentioned next as we will use these to discuss the finite precision properties of FOA based DMD.

Let $Y\in\mathbb{R}^{M\times N}$ be the matrix whose QR factorization is to be computed. Then QR factorization using CGS with reorthogonalization produces $\hat{Q}$ and $\hat{R}$ under the numerical nonsingularity of $Y$ \citep{giraud2005rounding} such that
\begin{equation}\label{eqn:qrfactbstab}
Y+\Delta Y=\hat{Q}\hat{R},\quad \|\Delta Y\|_F\le c_1\epsilon_m \|Y\|_F,
\end{equation}
\begin{equation}\label{eqn:qrfactorth}
\|I-\hat{Q}^H\hat{Q}\|_2\le c_2\epsilon_m.
\end{equation} 
Equation \ref{eqn:qrfactbstab} shows that computed factors $\hat{Q}$ and $\hat{R}$ are the exact QR factorization of a perturbed matrix bounded by the machine epsilon in a relative sense. This is called the backward error estimate. Equation \ref{eqn:qrfactorth} shows that the computed $\hat{Q}$ is orthonormal upto machine precision. Using the above estimates for QR decomposition \cite{giraud2005rounding} also derived backward error estimates for the standard Arnoldi method (used in Algorithm \ref{algo:rayleighritzarnoldi}) using the corresponding orthogonalization kernel. Even though the modified FOA is theoretically equivalent to the standard Arnoldi method, due to the unavailability of $A$ in FOA based DMD, the error estimates for backward error of computed quantities of the standard Arnoldi do not apply. However, the computed Arnoldi vectors using modified FOA will still be orthonormal upto machine precision.

To obtain the backward error estimate of FOA based DMD without rank truncation, we first decompose $A\hat{V}_1^{N-1}\hat{\beta}_{N-1}-\hat{V}_1^{N}\hat{\bar{H}}_N\hat{\beta}_{N-1}$ into different components which we can individually estimate as follows.
\begin{equation} \label{eqn:decompfoabaseddmd}
A\hat{V}_1^{N-1}\hat{\beta}_{N-1}-\hat{V}_1^{N}\hat{\bar{H}}_N\hat{\beta}_{N-1} = A\left(\hat{V}_1^{N-1}\hat{\beta}_{N-1}-X_1^{N-1}\right) + \left(X_2^N-\hat{V}_1^{N}\hat{\beta}_{1:N,2:N}\right)+\hat{V}_1^N\left(\hat{\beta}_{1:N,2:N}-\hat{\bar{H}}_N\hat{\beta}_{N-1}\right)
\end{equation}
We show in \ref{app:foabaseddmdberr} (Equation \ref{eqn:foabaseddmddecomperr}) that the error in computed factorization of $X_1^{N-1}$ into the orthonormal vectors $\hat{V}_1^{N-1}$ and upper triangular matrix $\hat{\beta}_{N-1}$ is of order $\epsilon_m\|X_1^{N-1}\|_2$ which is important, as it implies that we can reliably use the columns of $\hat{V}_1^{N-1}$ as basis vectors for the range of $X_1^{N-1}$. Also, we show in \ref{app:foabaseddmdberr} (Equation \ref{eqn:foabaseddmddecomperr}) that each term in the right hand side of Equation \ref{eqn:decompfoabaseddmd} is of order $\epsilon_m\|X_1^{N-1}\|_2$ i.e.,
\begin{equation} \label{eqn:foabaseddmdberrwithbeta}
\|A\hat{V}_1^{N-1}\hat{\beta}_{N-1}-\hat{V}_1^N\hat{\bar{H}}_N\hat{\beta}_{N-1}\|_F\le C_1\left(\|A\|_2,\|\hat{\bar{H}}_N\|_2,M,N\right)\epsilon_m\|X_1^{N-1}\|_2+O\left(\epsilon_m^2\right),
\end{equation}
where $C_1\left(\|A\|_2,\|\hat{\bar{H}}_N\|_2,M,N\right)$ is a constant which is a function of $\|A\|_2$, $\|\hat{\bar{H}}_N\|_2$, $M$ and $N$. The backward error of the Arnoldi relation using FOA based DMD can then be obtained as follows.
\begin{equation} \label{eqn:foabaseddmdberransatz}
\begin{split}
\|A\hat{V}_1^{N-1}-\hat{V}_1^N\hat{\bar{H}}_N\|_F &=\|\left(A\hat{V}_1^{N-1}\hat{\beta}_{N-1}-\hat{V}_1^N\hat{\bar{H}}_N\hat{\beta}_{N-1}\right)\hat{\beta}_{N-1}^{-1}\|_F, \\
&\le \|A\hat{V}_1^{N-1}\hat{\beta}_{N-1}-\hat{V}_1^N\hat{\bar{H}}_N\hat{\beta}_{N-1}\|_F\|\hat{\beta}_{N-1}^{-1}\|_2\left(\because \|AB\|_F\le\|A\|_F\|B\|_2\right).
\end{split}
\end{equation}
From Equation \ref{eqn:errorinfact}, we also see that $\|\hat{\beta}_{N-1}^{-1}\|_2\le\|X_1^{{N-1}^\dagger}\|_2+O\left(\epsilon_m\right)$, where $X_1^{{N-1}^\dagger}$ is the pseudoinverse of $X_1^{N-1}$. Using this in Equation \ref{eqn:foabaseddmdberransatz}, we get the following backward error estimates,
\begin{equation} \label{eqn:foabaseddmdberr}
\begin{split}
\|A\hat{V}_1^{N-1}-\hat{V}_1^N\hat{\bar{H}}_N\|_F\le& C_1\epsilon_m\kappa_2\left(X_1^{N-1}\right) + O\left(\epsilon_m^2\right),\\
\|\hat{V}_1^{{N-1}^H}A\hat{V}_1^{N-1}-\hat{H}_{N-1}\|_F\le& C_1\epsilon_m\kappa_2\left(X_1^{N-1}\right) + O\left(\epsilon_m^2\right),
\end{split}
\end{equation}
where $\kappa_2\left(X_1^{N-1}\right)$ is the 2-norm condition number defined as $\|X_1^{N-1}\|_2\|X_1^{{N-1}^{\dagger}}\|_2$.

The estimates obtained in Equation \ref{eqn:foabaseddmdberr} are important because they imply that the error in the computed projection $\||\hat{V}_1^{{N-1}^H}A\hat{V}_1^{N-1}-\hat{H}_{N-1}\|_F$ is of order $\epsilon_m\kappa_2\left(X_1^{N-1}\right)$. So, as the condition number of the snapshot matrix increases, the error in the computed projection due to finite precision arithmetic increases. As we will see later, this $\kappa_2\left(X_1^{N-1}\right)$ dependence in the computed projection is also true for SVD based DMD and sometimes it might also depend on higher powers of $\kappa_2\left(X_1^{N-1}\right)$ which is the case in streaming DMD. This dependence primarily arises since we are working with the snapshot vectors and not with the linear mapping $A$ itself, which is the case for the standard Arnoldi method (used in Algorithm \ref{algo:rayleighritzarnoldi}). So, DMD methods without rank truncation should be used to compute DMD modes and eigenvalues when $\kappa_2\left(X_1^{N-1}\right)\ll 1/\epsilon_m$. Next, we discuss the rationale behind FOA based DMD with rank truncation, to reduce the finite precision error in the computed projected matrix which can be used when $\kappa_2\left(X_1^{N-1}\right)\approx 1/\epsilon_m$.

From equation \ref{eqn:foabaseddmdberrwithbeta}, note that the error in $A\hat{V}_1^{N-1}\hat{\beta}_{N-1} - \hat{V}_1^N\hat{\bar{H}}_N\hat{\beta}_{N-1}$ is small and does not depend on condition number of snapshot matrix. It is in Equation \ref{eqn:foabaseddmdberransatz} when we multiply by inverse of $\hat{\beta}_{N-1}$ that we introduce dependence of $A\hat{V}_1^{N-1} - \hat{V}_1^N\hat{\bar{H}}_N$ on the condition number of snapshots. If the smallest singular value of $\hat{\beta}_{N-1}$ is very close to 0, then $\|\hat{\beta}^{-1}_{N-1}\|_2$ is very large thereby indicating large error in $A\hat{V}_1^{N-1} - \hat{V}_1^N\hat{\bar{H}}_N$ and $\hat{V}_1^{{N-1}^H}A\hat{V}_1^{N-1}-\hat{H}_{N-1}$. So, the computed projection of $A$ onto the Krylov subspace will have large error. However, we can choose a subspace of the Krylov subspace on which we can more accurately compute the projection of $A$.

Consider the SVD (in exact arithmetic) of the upper triangular matrix $\hat{\beta}_{N-1}$ i.e. $\hat{\beta}_{N-1}=U\Sigma W^H$. Since, $X_1^{N-1}-\hat{V}_1^{N-1}\hat{\beta}_{N-1}$ is of size $\epsilon_m\|X\|_2$, the SVD of $X_1^{N-1}$ is then
\begin{equation}
X_1^{N-1}=\hat{V}_1^{N-1}U\Sigma W^H+O(\epsilon_m\|X_1^{N-1}\|_2).
\end{equation}
Define $U_r$ as that first 'r' columns of matrix $U$, $W_r$ as the first 'r' columns of matrix $W$ and $\Sigma_r$ as the $r\times r$ sub-matrix of $\Sigma$.  Choosing the range of first $r$ singular vectors of $X_1^{N-1}$ as the subspace to perform Galerkin projection and $\hat{V}_1^{N-1}U_r$ as the corresponding basis (i.e. $V=Q=V_1^{N-1}U_r$ in Equation \ref{eqn:pdmdgalprob}), we can obtain the corresponding backward error as follows.
\begin{equation} \label{eqn:foabaseddmdrtkberr}
\begin{split}
\|A\hat{V}_1^{N-1}U_r-\hat{V}_1^N\hat{\bar{H}}_NU_r\|_F &=\|\left(A\hat{V}_1^{N-1}\hat{\beta}_{N-1}-\hat{V}_1^N\hat{\bar{H}}_N\hat{\beta}_{N-1}\right)\hat{\beta}_{N-1}^{-1}U_r\|_F, \\
&\le \|A\hat{V}_1^{N-1}\hat{\beta}_{N-1}-\hat{V}_1^N\hat{\bar{H}}_N\hat{\beta}_{N-1}\|_F\|\hat{\beta}_{N-1}^{-1}U_r\|_2\left(\because \|AB\|_F\le\|A\|_F\|B\|_2\right), \\
&= \|A\hat{V}_1^{N-1}\hat{\beta}_{N-1}-\hat{V}_1^N\hat{\bar{H}}_N\hat{\beta}_{N-1}\|_F\|\Sigma_r^{-1}\|_2, \\
\|{\hat{V}_1^{N-1}U_r}^HA\hat{V}_1^{N-1}U_r-U_r^H\hat{H}_{N-1}U_r\|_F&\le \|A\hat{V}_1^{N-1}U_r-\hat{V}_1^N\hat{\bar{H}}_NU_r\|_F(1+O(\epsilon_m)) \\
&\le C_1\epsilon_m\|X_1^{N-1}\|_2\|\Sigma_r^{-1}\|_2 + O(\epsilon_m^2)\text{ (Using Equation \ref{eqn:foabaseddmdberrwithbeta})}
\end{split}
\end{equation}
Here, we considered the SVD in exact arithmetic as the finite precision error in the computed SVD will only contribute to terms of order $\epsilon_m^2$.

The span of the columns of $\hat{V}_1^{N-1}U_r$ represent a subspace of the Krylov subspace (equation \ref{eq:krylovsubspace}). The computed projection of $A$ onto the subspace formed by the columns of $\hat{V}_1^{N-1}U_r$ is then $U_r^H\hat{H}_{N-1}U_r$. Since, $\|\Sigma_r^{-1}\|_2$ is much less than $\|X_1^{{N-1}^{\dagger}}\|_2$ (for a suitably chosen $r$), Equation \ref{eqn:foabaseddmdrtkberr} tells us that the error in the computed projection of $A$ onto subspace $\hat{V}_1^{N-1}U_r$ is smaller than that onto $\hat{V}_1^{N-1}$. This explains the rationale behind the rank truncation option of the proposed FOA based DMD algorithm shown in Algorithm \ref{algo:foabaseddmd}. Also, the DMD eigenvalue $\lambda$ of $U_r^H\hat{H}_{N-1}U_r$ and DMD eigenvector $V_1^{N-1}U_rz$, where $z$ is the eigenvector corresponding to $\lambda$, are nearly the same as the ones that one would obtain using DMD with rank truncated SVD \citep{schmid2010dynamic}.

\paragraph{SVD based DMD}
Next, we perform finite precision error analysis of SVD based DMD \citep{schmid2010dynamic} without rank truncation. The computed SVD of $X_1^{N-1}$ has the property \citep{golub2012matrix},
\begin{equation}
\begin{split}
&X_1^{N-1} + E = \left(\hat{U}+\delta\hat{U}\right)\hat{S}\left(\hat{W}+\delta\hat{W}\right)^H,\\
&\|E\|_2\le p_1\epsilon_m\|X_1^{N-1}\|_2,\\
&\|\delta\hat{U}\|_2\le p_2\epsilon_m,\quad \|\delta\hat{V}\|_2\le p_3\epsilon_m.
\end{split}
\end{equation}
where $p_1,p_2,p_3$ are polynomials of reasonable degree in the matrix dimensions of $X_1^{N-1}$. First, we estimate how close $A\hat{U}\hat{S}\hat{W}^H$ is to $X_2^N$.
\begin{equation} \label{eqn:svdbaseddmdfirststep}
\begin{split}
A\hat{U}\hat{S}\hat{W}^H&=A\left(X_1^{N-1}+E-\delta\hat{U}\hat{S}\delta\hat{W}^H - \hat{U}\hat{S}\delta\hat{W}^H\right)\\
&=X_2^N+A\left(E-\delta\hat{U}\hat{S}(\hat{W}+\delta\hat{W})^H - \hat{U}\hat{S}\delta\hat{W}^H\right).
\end{split}
\end{equation}
Multiplying the above equation by $\hat{W}\hat{S}^{-1}$ from the right, we have 
\begin{equation}
A\hat{U}=X_2^N\hat{W}\hat{S}^{-1} + A\left(E-\delta\hat{U}\hat{S}(\hat{W}+\delta\hat{W})^H - \hat{U}\hat{S}\delta\hat{W}^H\right)\hat{W}\hat{S}^{-1}.
\end{equation}
Multiplying the above equation by $U^H$ on left and accounting for the error in matrix-matrix multiplication,
\begin{equation} \label{eqn:svdbaseddmdberrnorkt}
\begin{split}
\hat{U}^HA\hat{U}-fl(\hat{U}^HX_2^N\hat{W}\hat{S}^{-1})=\left[\hat{U}^HX_2^N\hat{W}\hat{S}^{-1}-fl(\hat{U}^HX_2^N\hat{W}\hat{S}^{-1})\right] + \\
\left[\hat{U}^HA\left(E-\delta\hat{U}\hat{S}(\hat{W}+\delta\hat{W})^H - \hat{U}\hat{S}\delta\hat{W}^H\right)\hat{W}\hat{S}^{-1}\right].
\end{split}
\end{equation}
Observe that each of the two bracketed terms in right hand side is of size $O(\epsilon_m\kappa_2(X_1^{N-1}))$. So, the error in the computed projection of $A$ is $O(\epsilon_m\kappa_2(X_1^N))$ even using the SVD approach. 

To obtain the finite precision error of rank truncated SVD, multiplying the Equation \ref{eqn:svdbaseddmdfirststep} by $\hat{W}_r\hat{S_r}^{-1}$, where $r$ is the chosen truncated rank, $\hat{W}_r$ is the first $r$ columns of $\hat{W}$ and $\hat{S}_r$ is the top left $r\times r$ submatrix of $\hat{S}$, we obtain
\begin{equation}
A\hat{U}_r = X_2^N\hat{W}_r\hat{S}_r^{-1} + A\left(E-\delta\hat{U}\hat{S}(\hat{W}+\delta\hat{W})^H - \hat{U}\hat{S}\delta\hat{W}^H\right)\hat{W}_r\hat{S}_r^{-1}.
\end{equation}
Similar to the analysis of SVD based DMD, multiplying the above equation by $U_r^H$ on left and accounting for the error in matrix-matrix multiplication,
\begin{equation}
\hat{U}_r^HA\hat{U}_r-fl(\hat{U}_r^HX_2^N\hat{W}_r\hat{S}_r^{-1})=\left[\hat{U}_r^HX_2^N\hat{W}_r\hat{S}_r^{-1}-fl(\hat{U}_r^HX_2^N\hat{W}_r\hat{S}_r^{-1})\right] + \\
\left[\hat{U}_r^HA\left(E-\delta\hat{U}\hat{S}(\hat{W}+\delta\hat{W})^H - \hat{U}\hat{S}\delta\hat{W}^H\right)\hat{W}_r\hat{S}_r^{-1}\right].
\end{equation}
We can see from the above equation that size of the 2 bracketed terms in right hand side is $\epsilon_m\|X_1^{N-1}\|_2\|\hat{S}_r^{-1}\|_2$ which is small when compared to the $O(\epsilon_m\kappa_2\left(X_1^{N-1}\right))$ in Equation \ref{eqn:svdbaseddmdberrnorkt} for a suitably chosen $r$. Hence, the finite precision error in the computed projected matrix for SVD based DMD with rank truncation is smaller than that without rank truncation.

\paragraph{Streaming DMD}
Streaming DMD \citep{hemati2014dynamic} uses QR decomposition to compute the basis vectors for the Krylov subspace (Equation \ref{eq:krylovsubspace}). We do not consider compression and derive the finite precision error for the computed projection of the linear mapping.
\begin{equation} \label{eqn:streamingdmdqrdecomp}
\begin{split}
AX_1^{N-1}&=X_2^N,\\
A\left(\hat{Q}_X\hat{R}_X+E_X\right) &= \left(\hat{Q}_Y\hat{R}_Y+E_Y\right),
\end{split}
\end{equation}
where $\hat{Q}_X,\hat{R}_X$ and $\hat{Q}_Y,\hat{R}_Y$ is the computed QR decomposition of $X_1^{N-1}$ and $X_2^N$ respectively. From Equation \ref{eqn:qrfactbstab}, it follows that the backward errors $E_X$ and $E_Y$, in the computed factors are bounded by $c_1\epsilon_m\|X_1^{N-1}\|_2$ and $c_1\epsilon_m\|X_2^N\|_2$ respectively. In streaming DMD, the projected linear mapping is computed as $\hat{Q}_X^H\hat{Q}_Y\left(\hat{R}_X\hat{R}_Y^H\right)fl\left(\hat{R}_X\hat{R}_X^H\right)^{-1}$ using an incremental method to compute the matrix products $\hat{R}_X\hat{R}_Y^H$ and $\hat{R}_X\hat{R}_X^H$. The finite precision arithmetic error $\|\hat{Q}_X^HA\hat{Q}_X-\hat{Q}_X^H\hat{Q}_Y\left(\hat{R}_Y\hat{R}_X^H\right)fl\left(\hat{R}_X\hat{R}_X^H\right)^{-1}\|_F$ can be computed as follows. Multiplying Equation \ref{eqn:streamingdmdqrdecomp} by $\hat{R}_X^H$ from the right and rearranging, we get
\begin{equation}
A\hat{Q}_X\hat{R}_X\hat{R}_X^H = \hat{Q}_Y\hat{R}_Y\hat{R}_X^H - AE_X\hat{R}_X^H + E_Y\hat{R}_X^H.
\end{equation}
Multiplying by  $\left(\hat{R}_X\hat{R}_X^H\right)^{-1}$ on the right and $\hat{Q}_X^H$ from the left we have,
\begin{equation}
  \begin{split}
    \hat{Q}_X^HA\hat{Q}_X-\hat{Q}_X^H\hat{Q}_Y\hat{R}_Y\hat{R}_X^H\left(\hat{R}_X\hat{R}_X^H\right)^{-1} &= \hat{Q}_X^H\left(E_Y\hat{R}_X^H-AE_X\hat{R}_X^H\right)\left(\hat{R}_X\hat{R}_X^H\right)^{-1}, \\
    \hat{Q}_X^HA\hat{Q}_X-\hat{Q}_X^H\hat{Q}_Y\hat{R}_Y\hat{R}_X^Hfl\left(\hat{R}_X\hat{R}_X^H\right)^{-1} &= \hat{Q}_X^H\left(E_Y\hat{R}_X^H-AE_X\hat{R}_X^H\right)\left(\hat{R}_X\hat{R}_X^H\right)^{-1}-\hat{Q}_X^H\hat{Q}_Y\hat{R}_Y\hat{R}_X^H\left(fl\left(\hat{R}_X\hat{R}_X^H\right)^{-1}-\left(\hat{R}_X\hat{R}_X^H\right)^{-1}\right).
   \end{split}
 \end{equation}
Observe that the size of right hand side in the above equation can be as large as order $\epsilon_m\left(\kappa_2(X_1^{N-1})\right)^2$. This is in contrast to FOA based and SVD based DMD without rank truncation whose finite precision arithmetic error in the computed projection is only order $\epsilon_m\kappa_2(X_1^{N-1})$. So, for a fixed condition number of snapshots, the finite precision error in the computed projection from streaming DMD is higher than FOA based and SVD based DMD. This is primarily because streaming DMD involves inversion of the matrix product $\hat{R}_X\hat{R}_X^H$. 


\begin{figure}[]
\centering
\includegraphics[width=.5\linewidth]{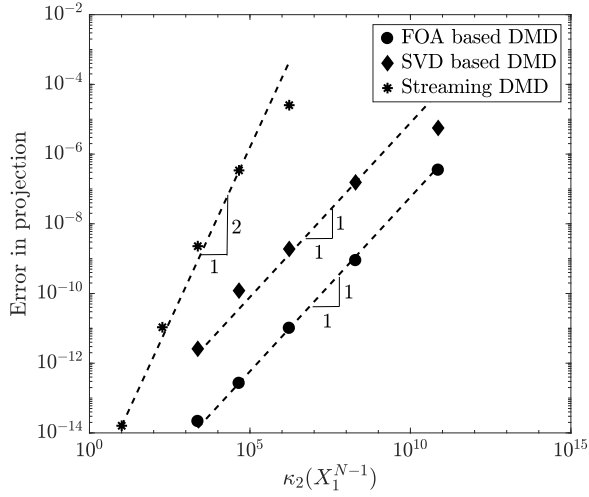}
\caption{Finite precision error in computed projection using different projected DMD methods.}
\label{fig:errprjdmd}
\end{figure}

Figure \ref{fig:errprjdmd} shows numerical evidence for the above obtained error estimates for different projected DMD methods without rank truncation. The linear mapping $A$ is chosen to be a Vandermonde matrix $\in\mathbb{R}^{50\times 50}$ generated in Matlab using $vander(linspace(0,1,50))$. Snapshots are generated by the application of A onto a random starting vector. The number of snapshots is varied to obtain a range of condition numbers of the snapshot matrix. The finite precision error in the computed projection is plotted as a function of the condition number of the snapshot matrix for each of the three projected DMD methods in Figure \ref{fig:errprjdmd}. As we can see from the figure, error in the computed projection using streaming DMD can depend on $\left(\kappa_2(X_1^{N-1})\right)^2$ whereas that from FOA based and SVD based DMD depends on $\kappa_2(X_1^{N-1})$. 

\subsubsection{Error indicators for DMD modes and eigenvalues} \label{sec:foabaseddmderrorind}

DMD computes approximate eigenvector and eigenvalues of the assumed linear mapping $A$. In exact arithmetic, using matrix $\bar{H}_N$ and the eigenvector $z_i$ it is possible to obtain error estimates associated with DMD eigenvalues and DMD eigenvectors \citep{saad2011numerical} in $\ell^2$ norm computed using FOA based DMD by using the relation
    \begin{equation} \label{eq:errordmdvec}
    \|AV_1^{N-1}z_i - \lambda_iV_1^{N-1}z_i\|_2 = \|h_{N,N-1}v_Ne^H_{N-1}z_i\|_2.
    \end{equation}
By monitoring the value of $\|h_{N,N-1}v_Ne^H_{N-1}z_i\|_2$ it is possible to monitor the accuracy of DMD eigenvector and eigenvalue pairs. This can be reasoned by the Bauer-Fike theorem \citep{saad2011numerical} which states that suppose $\left(\lambda_a,v_a\right)$ is an approximate eigenvalue-eigenvector pair, then the error in $\lambda_a$ is related to the residual $\|Av_a-\lambda_av_a\|_2$ such that
\begin{equation}
|\lambda_a-\lambda|\le\kappa_2(V) \frac{\|Av_a-\lambda_a v_a\|_2}{\|v_a\|_2}
\end{equation}
where $V$ is the matrix of right eigenvectors of $A$ (assuming that its eigenvectors are linearly independent). So, if the DMD eigenvector has unit magnitude, the error in DMD eigenvalue $\lambda_i$ computed using FOA based DMD is indicated by $\|h_{N,N-1}v_Ne^H_{N-1}z_i\|_2$. This can be used to devise stopping criterions for acquiring new snapshots when using FOA based DMD.

Suppose, we use FOA based DMD with rank truncation, the error indicator for the DMD eigenvectors and DMD eigenvalues can be obtained as follows. In exact arithmetic, 
\begin{equation}
\begin{split}
AV_1^{N-1}U_rz_i - \lambda_i V_1^{N-1}U_rz_i &= V_1^{N-1}(I-U_rU_r^H)H_{N-1}U_rz_i + h_{N,N-1}v_Ne_{N-1}^HU_rz_i, \\
\|AV_1^{N-1}U_rz_i - \lambda_i V_1^{N-1}U_rz_i\|_2&= \|V_1^{N-1}(I-U_rU_r^H)H_{N-1}U_rz_i + h_{N,N-1}v_Ne_{N-1}^HU_rz_i\|_2,\\
\|AV_1^{N-1}U_rz_i - \lambda_i V_1^{N-1}U_rz_i\|_2&= \|\begin{bmatrix}(I-U_rU_r^H)H_{N-1}\\h_{N,N-1}e_{N-1}^H\end{bmatrix}U_rz_i\|_2.
\end{split}
\end{equation}
The error in DMD eigenvector $V_1^{N-1}U_rz_i$ (unit magnitude in $\ell^2$ norm) and eigenvalue $\lambda_i$ pair computed using FOA based DMD with rank truncation is indicated by $\|\begin{bmatrix}(I-U_rU_r^H)H_{N-1}\\h_{N,N-1}e_{N-1}^H\end{bmatrix}U_rz_i\|_2$.

Next, we investigate how close the computed error indicator for the DMD eigenvector and eigenvalue obtained using FOA based DMD without rank truncation is to the actual error using the finite precision results derived in previous section. From Equation \ref{eqn:foabaseddmdberrwithbeta} and assuming $\hat{\beta}_{N-1}$ to be full rank, we have
\begin{align}
\|A\hat{V}_1^{N-1}\hat{\beta}_{N-1}y_i - \hat{V}_1^N\hat{\bar{H}}_N\hat{\beta}_{N-1}y_i\|_2 &\le C_1\epsilon_m\|X_1^{N-1}\|_2\|y_i\|_2, \\
\hat{\beta}_{N-1}y_i&=z_i \label{eqn:definey},
\end{align}
where $\lambda_i,z_i$ is an exact eigenvalue eigenvector pair of $\hat{H}_{N-1}$. Here, we assume for simplicity that the $y_i$ is the exact solution of $\hat{\beta}_{N-1}y_i=z_i$ and $z_i$ is the exact eigenvector of $\hat{H}_{N-1}$. Then, the accuracy of the error indicator is 
\begin{align}
\|A\hat{V}_1^{N-1}z_i - \lambda_i\hat{V}_1^{N-1}z_i - \hat{h}_{N,N-1}\hat{v}_Ne_{N-1}^Hz_i\|_2 \le C_1\epsilon_m\|X_1^{N-1}\|_2\|y_i\|_2.
\end{align}
The above equation tells us that the computed error indicator $\hat{h}_{N,N-1}\hat{v}_Ne_{N-1}^Hz_i$ is a good estimate of the actual error $A\hat{V}_1^{N-1}z_i - \lambda_i\hat{V}_1^{N-1}z_i$ if the magnitude of $y_i$ obtained from solving Equation \ref{eqn:definey} is small. This is essentially the same as saying that the component of $z_i$ along the left singular vectors corresponding to near zero singular values of $\hat{\beta}_{N-1}$ is small.

However, for FOA based DMD with rank truncation, the computed error indicator $\|\begin{bmatrix}(I-U_rU_r^H)H_{N-1}\\h_{N,N-1}e_{N-1}^H\end{bmatrix}U_rz\|_2$ for the computed DMD eigenvector and eigenvalue will be very close to the actual error for a suitably chosen truncated rank $r$.  This is because the backward error in Equation \ref{eqn:foabaseddmdrtkberr} is small for appropriately chosen $r$ and is controlled by $\|\Sigma_r^{-1}\|_2$ rather than $\|\hat{\beta}_{N-1}^{-1}\|_2$. Numerical experiments shown in section \ref{sec:linchanflowsnaps} support the above arguments regarding finite precision quality of the FOA based DMD error indicators.

\subsubsection{Reconstruction error and interpolation arguments} \label{sec:reconerrorandinterp}

Let $\varphi_j:=V_1^{N-1}z_j;j=1,\dots,N-1$. Since, we have assumed that the DMD eigenvectors are linearly independent, $span\{\varphi_i\}_{i=1}^{N-1}$ is same as $K_{N-1}(A,\psi_1)$. So, we should be able to reconstruct atleast the first $N-1$ snapshots using $\{\varphi\}_{j=1}^{N-1}$ exactly \citep{rowley2009spectral}.

\cite{rowley2009spectral} showed that all snapshot vectors except the last one in the sequence can be exactly reconstructed from the DMD modes. The reconstruction error of the last snapshot was shown to be that of its orthogonal projection onto the previous ones. Here, we obtain the same results using the Arnoldi basis vectors in the context of FOA based DMD. Suppose $\psi=\sum_{j=1}^{N-1}c_j\varphi_j$. Then, say we approximate $A\psi$ as $\sum_{j=1}^{N-1}\lambda_jc_j\varphi_j$. The error involved in this approximation can be obtained as follows
    \begin{equation}
    \begin{split}
        &A\psi - \sum_{j=1}^{N-1}\lambda_j c_j \varphi_j = \sum_{j=1}^{N-1}c_j\left(A\varphi_j - \lambda_j \varphi_j\right) \\
        &= \sum_{j=1}^{N-1}c_j\left(h_{N,N-1}v_Ne_{N-1}^Hz_j\right) = h_{N,N-1}\left(\sum_{j=1}^{N-1}c_je_{N-1}^Hz_j\right)v_N,
    \end{split}
    \end{equation}
    \begin{equation}\label{eq:reconerror}
            \|A\psi - \sum_{j=1}^{N-1}\lambda_j c_j \varphi_j\|_2 = |h_{N,N-1}||\sum_{j=1}^{N-1}c_je_{N-1}^Hz_j|.
    \end{equation}
    $\psi=\sum_{j=1}^{N-1}c_j\varphi_j$ can be written in matrix form as $\psi=[\varphi]\{c\}$ where $[\varphi]$ is matrix formed by stacking $\varphi_j$ as columns and $\{c\}$ is the vector of coefficients. i.e. $[\varphi] = [\varphi_1\quad\dots\quad\varphi_{N-1}]$ and $\{c\}=[c_1\quad\dots\quad c_{N-1}]^H$. Let $[z]:=[z_1\quad\dots\quad z_{N-1}]$. Then, we have 
    \begin{equation}
    \begin{split}
        \psi &= V_1^{N-1}[z]\{c\},\\
        \psi &= V_1^{N-1}\{\tilde{c}\}\quad(\{\tilde{c}\}:=[z]\{c\}).
    \end{split}
    \end{equation}
Suppose $h_{N,N-1}$ is not 0. Then, $\|A\psi - \sum_{j=1}^{N-1}\lambda_j c_j \varphi_j\|_2$ is 0 if $\sum_{j=1}^{N-1}c_je_{N-1}^Hz_j$ is 0. i.e. the last entry in the vector $\{\tilde{c}\}$ should be 0 or equivalently, $\psi=V_1^{N-2}\{\bar{c}\}$ where $\{\bar{c}\}=[\tilde{c}_1\quad\dots\quad\tilde{c}_{N-2}]^T$. Therefore, the approximation $\sum_{j=1}^{N-1}\lambda_jc_j\varphi_j$ to $A\psi$ is exact if atleast one of the below conditions are satisfied.
    \begin{enumerate}
        \item $\psi$ is in the range of $V_{N-2}$.
        \item $h_{N,N-1}=0$. i.e. if the last snapshot can be written as linear combination of the previous snapshots.
    \end{enumerate}
The above 2 conditions are essentially consequences of the fact that we use Galerkin projection onto the subspace $K_{N-1}(A,\psi_1)$.

The FOA based DMD method stops if linearly dependent snapshots are present. i.e. say $\psi_{j+1}$ can be expressed as a linear combination of $\{\psi_1,\dots,\psi_j\}$. Then, in the $j^{th}$ step of Arnoldi method, $h_{j+1,j}$ is 0. The Galerkin method with the previously computed $v_i$'s will give exact eigenvectors and eigenvalues associated with the linear mapping $A$ that connects the snapshot vectors.

In a continuous sense, DMD generates an interpolant through the snapshot vectors. The interpolant can be defined as
\begin{equation}
\begin{split}
I^N\psi(t) := \sum_{j=1}^{N-1}e^{\omega_j t} d_j\varphi_j, \\
I^N\psi_k:=I^N\psi(t_k),
\end{split}
\end{equation}
where $\omega_j:=ln(\lambda_j)/\Delta t$, $\Delta t$ is the time spacing between the snapshots, $t_k:=(k-1)\Delta t$ and $\psi_1=\sum_{j=1}^{N-1}d_j\varphi_j$.
    Equation \ref{eq:reconerror} implies that snapshots $\psi_1$ to $\psi_{N-1}$ can always be exactly reconstructed from the DMD eigenmodes and DMD eigenvalues. i.e.
\begin{equation}
\psi_k=I^N\psi_k\quad k=1,\dots,N-1
\end{equation}
and the error associated with reconstruction of $I^N\psi_{N-1}$ is
\begin{equation}
\|I^N\psi_{N}-\psi_N\|_2 = |h_{N,N-1}||\sum_{j=1}^{N-1}d_j\lambda_j^{N-2}e_{N-1}^Hz_j|.
\end{equation}

In the linear case, the interpolant $I^N\psi(t)$ approximates the evolution $e^{Bt}\psi_1$, where $A=e^{B\Delta t}$. For the nonlinear case, suppose we use $P$ snapshots in the FOA based DMD algorithm starting from snapshot $\psi_1$ and $h_{P,P-1}$ turns to be 0, then DMD eigenvectors and eigenvalues are the exact eigenpairs of $A$. All $P$ snapshots can be constructed exactly using DMD eigenvectors and DMD eigenvalues and consequentially the interpolant $I^P\psi(t)$. Now, if we use $I^P\psi(t)$ to predict the observable vectors at future times, the error in the prediction is an indication of how close the span of observables is to a Koopman function and eigenvalues \citep{tu2014dynamic} of the system. Numerical experiments on evaluating the quality of approximate Koopman eigenfunctions obtained from DMD for different choices of set of observables is carried out in \cite{zhang2017evaluating}.

The coefficients $\{c\}$ of first snapshot $\psi_1$ in the basis formed by DMD eigenmodes can be obtained by solving the following matrix problem of size $(N-1)\times(N-1)$.
    \begin{equation} \label{eq:exrecon}
    \begin{split}
        [\varphi]\{c\} &= \psi_1,\\
        [z]\{c\} &= \|\psi_1\|_2e_1.
    \end{split}
    \end{equation}
If rank truncated FOADMD is used, then the coefficients $\{c\}$ of the first snapshot in eigenvector basis such that $\|\psi_i-[V_1^{N-1}U_r][z]\{c\}\|_2$ is minimized, can be obtained by solving
    \begin{equation} \label{eq:exreconwrt}
    \begin{split}
        [z]\{c\} &= \|\psi_1\|_2U_r^He_1.
    \end{split}
    \end{equation}

\subsubsection{Parallel scaling} \label {sec:parallelscaling}

\begin{figure}[]
\centering
\begin{subfigure}{.5\textwidth}
  \centering
  \includegraphics[width=\linewidth]{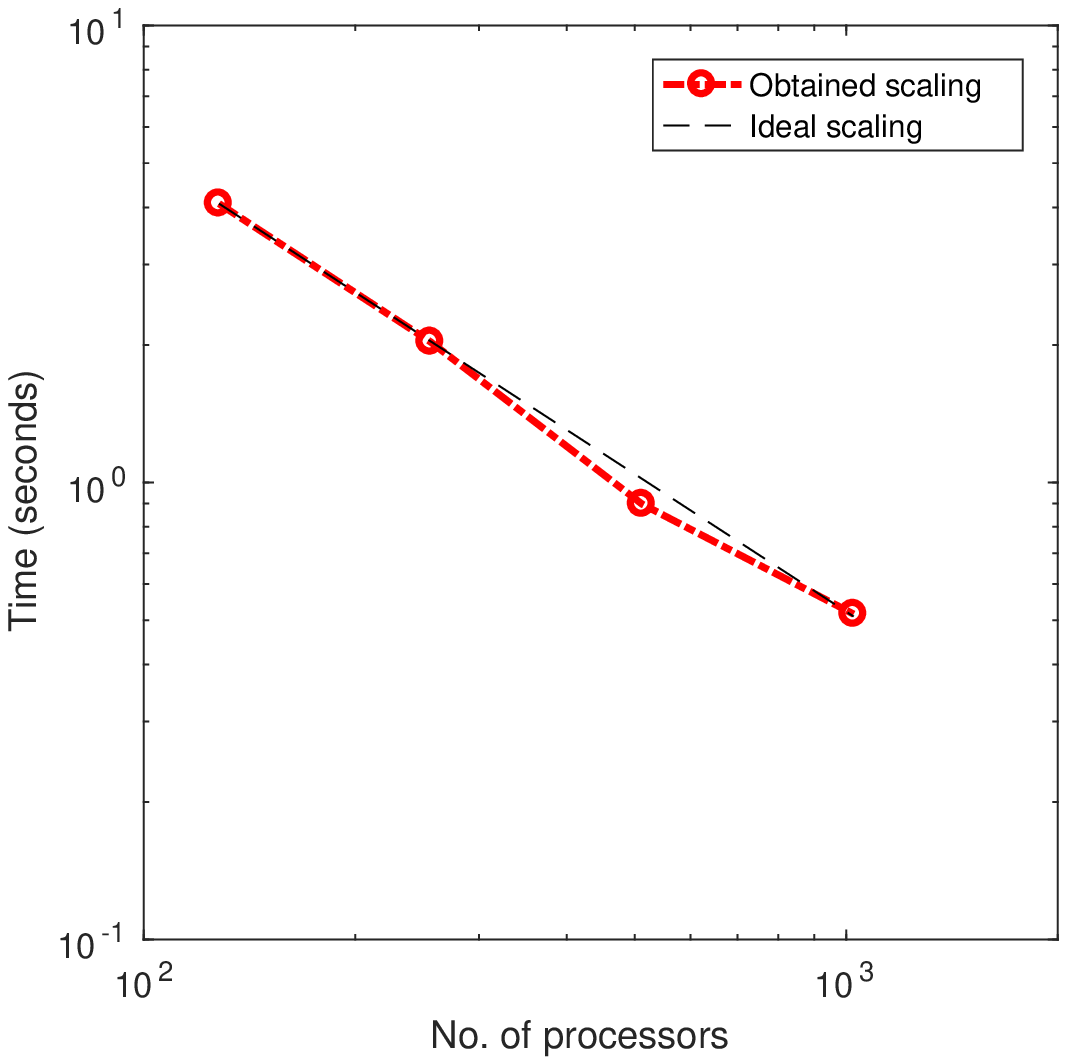}
  \caption{Strong scaling on TACC Stampede2 cluster (8 million, 101 snapshots).}
  \label{fig:itascascaling}
\end{subfigure}%
\begin{subfigure}{.5\textwidth}
  \centering
  \includegraphics[width=\linewidth]{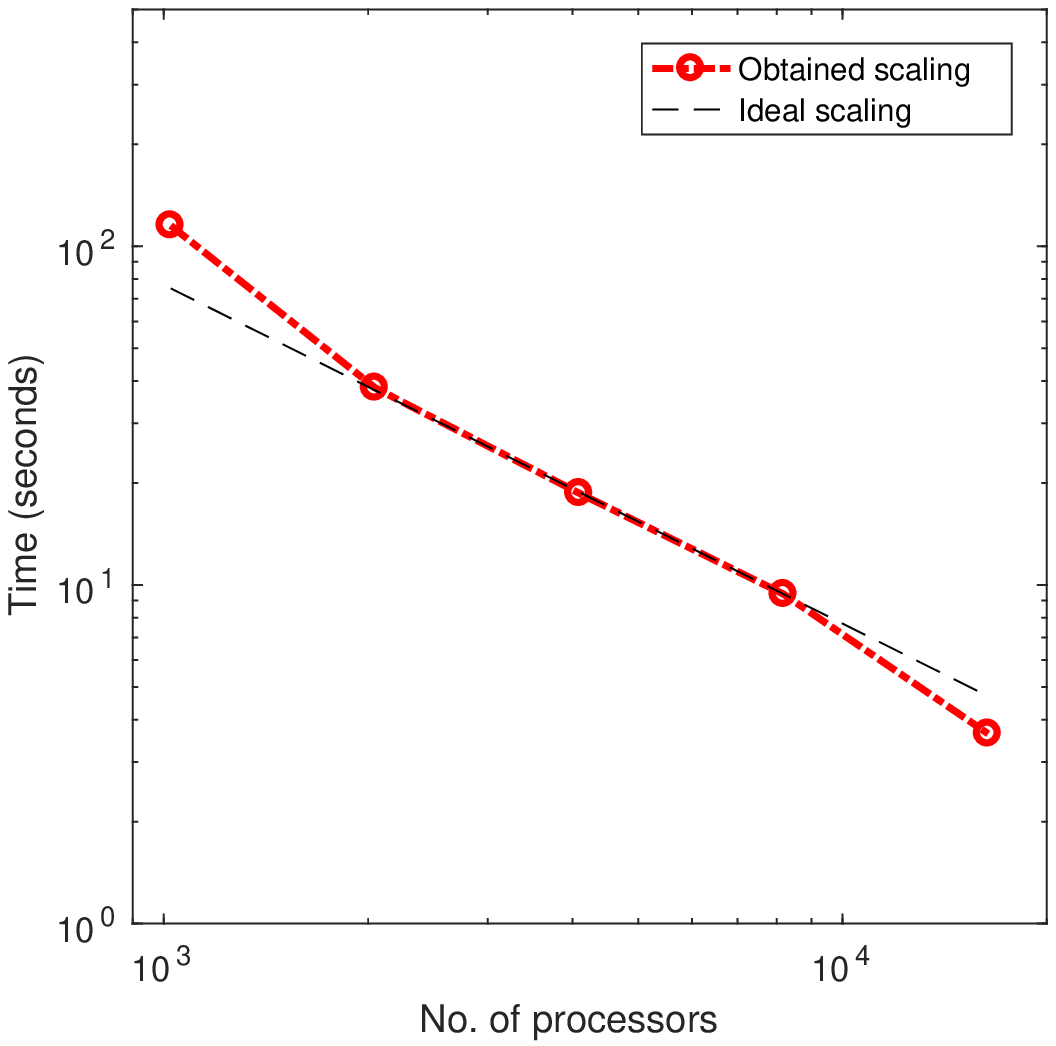}
  \caption{Strong scaling on TACC Stampede2 cluster (240 million, 201 snapshots).}
  \label{fig:stampede2scaling}
\end{subfigure}
\caption{Strong scaling of $H_{N-1}$ computation using FOA based DMD (Algorithm. \ref{algo:foabaseddmd}).}
\end{figure}

The computational kernels of FOA based DMD algorithm include vector additions, dot products and matrix vector multiplications which can be parallelized. The snapshot data can be partitioned row-wise among different processors. Figs. \ref{fig:itascascaling} and \ref{fig:stampede2scaling} show strong scaling for the computation of projected matrix $H_{N-1}$ using FOA based DMD algorithm on Texas Advanced Computing Center (TACC) Stampede2 Knights Landing cluster upto 16384 processors respectively. Both plots show time taken to generate Arnoldi vectors and Hessenberg matrix for different number of processors. The strong scaling shown in Fig. \ref{fig:itascascaling} utilized 101 snapshots of size 8 million each and in Fig. \ref{fig:stampede2scaling} used 201 snapshots of size 240 million each. The snapshot matrix was filled with random numbers as the operation count and scaling properties of the algorithm is independent of the content of the snapshot matrix. Fig. \ref{fig:itascascaling} and \ref{fig:stampede2scaling} show good scaling properties of the algorithm. Since we use CGS with reorthogonalization, $O(N)$ number of messages are exchanged between processors. This is in contrast to MGS orthogonalization kernel which would require $O(N^2)$ number of messages.

\section{Numerical experiments} \label{sec:numericalexperiments}

We illustrate application of FOA based DMD (Algorithm \ref{algo:foabaseddmd}) on three problems, i) snapshots from a linearized channel flow simulation, ii) snapshots of vorticity field from flow over a circular cylinder and iii) snapshots from turbulent flow simulation of jets in cross flow.

\subsection{Linearized channel flow simulation at Re=10000} \label{sec:linchanflowsnaps}
\begin{figure}[]
  \centering
  \includegraphics[width=0.4\textwidth]{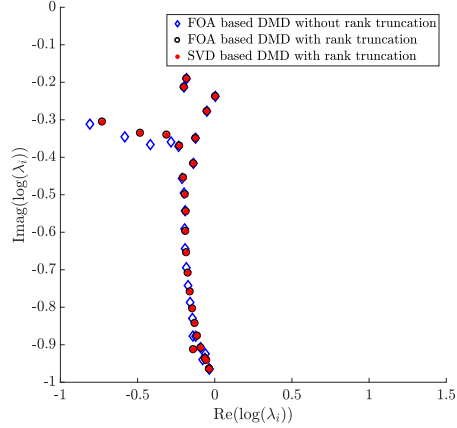}
  \caption{Comparison of DMD eigenvalues for linearized channel flow simulation snapshots.}
  \label{fig:cheigcomp}
\end{figure}

The dataset used to perform DMD analysis of linearized channel flow simulation was obtained from \cite{jovanovic2014sparsity}. For a description of the numerical method and mesh resolution refer \cite{jovanovic2014sparsity}. This dataset is extremely ill-conditioned. The condition number of snapshot matrix is $\approx 10^{17}$. Even though, 100 snapshots are present, the numerical rank of the dataset is 26. We define numerical rank as the number of singular values of snapshot matrix that are larger than $N*eps(\|A\|_2)$, where $eps(x)$ is the distance between $|x|$ and the next larger double precision floating point number. Also, the linear mapping $A$ is available. So, we can compute the error involved in projection of $A$, DMD eigenvectors and eigenvalues and assess quality of proposed error indicators when the snapshots become extremely ill-conditioned. From the same set of snapshots as used in \cite{jovanovic2014sparsity} we compute the DMD eigenvectors and eigenvalues and compare to the SVD based DMD. Streaming DMD without compression would not yield reliable results for this case due to the very high condition number of snapshot matrix. The size of each snapshot is 150 and a total of 100 snapshots are used. DMD eigenvalues obtained using FOA based DMD with and without rank truncation is compared to SVD based DMD in figure \ref{fig:cheigcomp}. The DMD eigenvectors and eigenvalues are ordered based on the values of coefficients obtained by projection of the first snapshot onto the DMD eigenmodes using equation \ref{eq:exrecon}.

Table \ref{tab:errorinprojection} shows the error in computed projection of $A$. The exact projection $P_A$ is calculated as $fl(Q^HAQ)$. Since, matrix-matrix multiplication with orthogonal matrix is backward-stable \citep{higham2002accuracy}, error in evaluation of $P_A$ will be very small. Finite precision error analysis of SVD based DMD and FOA based DMD without rank truncation showed that the error in computed projection is $O(\kappa_2(X_1^N)\epsilon_m)$. This is consistent with the numerical results shown in table \ref{tab:errorinprojection}.  Also, the large error associated with streaming DMD is because of its dependence on higher powers of $\kappa_2(X_1^{N-1})$ as explained in section \ref{sec:fpefoabaseddmd}. The increase in accuracy of computed projection with rank truncated versions of SVD based DMD and FOA based DMD seen in Table \ref{tab:errorinprojection} is explained in section \ref{sec:fpefoabaseddmd}.
\begin{table}
\begin{center}
\begin{tabular}{| c | l | l | l |}
\hline
$\|P_A-\hat{P}_A\|_2$& FOA based DMD & SVD based DMD & Streaming DMD \\ \hline
Without rank truncation &2.22e+01 &6.03+00 & 1.05e+09 \\
With rank truncation &5.77e-04 &2.09e-03 &- \\ \hline
\end{tabular}
\end{center}
\caption{Error $\|P_A-\hat{P}_A\|_2$ in computed projection of $A$ for FOA based DMD, SVD based DMD and streaming DMD for linearized channel flow case. $P_A$ is the exact projection and $\hat{P}_A$ is the computed projection.}
\label{tab:errorinprojection}
\end{table}

Next, we consider the accuracy of computed DMD eigenvectors and eigenvalues using FOA based DMD with and without rank truncation. Tables \ref{tab:errorcompnotrunc} and \ref{tab:errorcomptrunc} show the comparison of actual error (Error$_a$) in eigenvalue-eigenvector pair to the prediction using the error indicator (Error$_p$) with and without rank truncation respectively for the 8 DMD modes with least error. The actual error and predicted error is defined as $\|AV_1^{N-1}z-\lambda V_1^{N-1}z\|_2$ and $|h_{N,N-1}e_{N-1}^Hz|$ respectively for the case without rank truncation and for the case with rank truncation, they are defined as $\|AV_1^{N-1}U_rz-\lambda V_1^{N-1}U_rz\|_2$ and $\|V_1^{N-1}\left(I-U_rU_r^H\right)H_{N-1}U_rz+h_{N,N-1}v_Ne_{N-1}^HU_rz\|_2$ respectively, where $\lambda$ and $z$ are the eigenvalue and eigenvector pair of the projected problem. As, we can see from table \ref{tab:errorcompnotrunc} and \ref{tab:errorcomptrunc}, the error in dominant eigenvalues and corresponding eigenvectors are of the same order of magnitude. However, the predicted error is more close to the actual error for the case with rank truncation than without rank truncation. This is because of the smaller backward error for the Arnoldi relation (equation \ref{eqn:foabaseddmdrtkberr}) for the case with rank truncation.

\begin{table}
\begin{minipage}{0.5\textwidth}
\centering
\begin{tabular}{|c|l|l|}
\hline
Eigenvalue & Error$_a$ & Error$_{p}$ \\
\hline
9.76e-01+(-2.36e-01)i & 3.75e-14 & 5.37e-15 \\ 
9.14e-01+(-2.60e-01)i & 1.07e-10 & 1.62e-11 \\ 
8.30e-01+(-3.02e-01)i & 5.79e-08 & 9.44e-09 \\ 
8.18e-01+(-1.57e-01)i & 1.90e-06 & 2.69e-07 \\ 
5.62e-01+(-7.19e-01)i & 1.39e-05 & 2.53e-06 \\ 
5.56e-01+(-7.56e-01)i & 8.03e-06 & 1.44e-06 \\ 
5.50e-01+(-7.93e-01)i & 5.32e-06 & 9.38e-07 \\ 
7.97e-01+(-3.52e-01)i & 3.77e-06 & 6.44e-07 \\ 
%
%
%
%
%

%
\hline
\end{tabular}
\caption{Comparison of actual error $\|AV_1^{N-1}z-\lambda V_1^{N-1}z\|_2$ with the predicted error without rank truncation.}
\label{tab:errorcompnotrunc}
\end{minipage}
\hfill
\begin{minipage}{0.5\textwidth}
\centering
\begin{tabular}{|c|l|l|}
\hline
Eigenvalue & Error$_a$ & Error$_p$ \\ \hline
9.76e-01+(-2.36e-01)i & 3.40e-13 & 3.40e-13 \\ 
9.14e-01+(-2.60e-01)i & 6.01e-10 & 6.00e-10 \\ 
8.30e-01+(-3.02e-01)i & 1.53e-07 & 1.53e-07 \\ 
8.18e-01+(-1.57e-01)i & 6.32e-06 & 6.32e-06 \\ 
5.62e-01+(-7.20e-01)i & 1.30e-05 & 1.30e-05 \\ 
5.57e-01+(-7.56e-01)i & 5.66e-06 & 5.66e-06 \\ 
5.50e-01+(-7.93e-01)i & 3.02e-04 & 3.02e-04 \\ 
7.97e-01+(-3.52e-01)i & 7.58e-06 & 7.57e-06 \\
%

%
%
%
%
%
\hline
\end{tabular}
\caption{Comparison of actual error $\|AV_1^{N-1}U_rz-\lambda V_1^{N-1}U_rz\|_2$ with the predicted error with rank truncation.}
\label{tab:errorcomptrunc}
\end{minipage}
\end{table}
\subsection{Flow over cylinder at Re=100}
\begin{figure}[]
\centering
\begin{subfigure}{.5\textwidth}
  \centering
  \includegraphics[width=\linewidth]{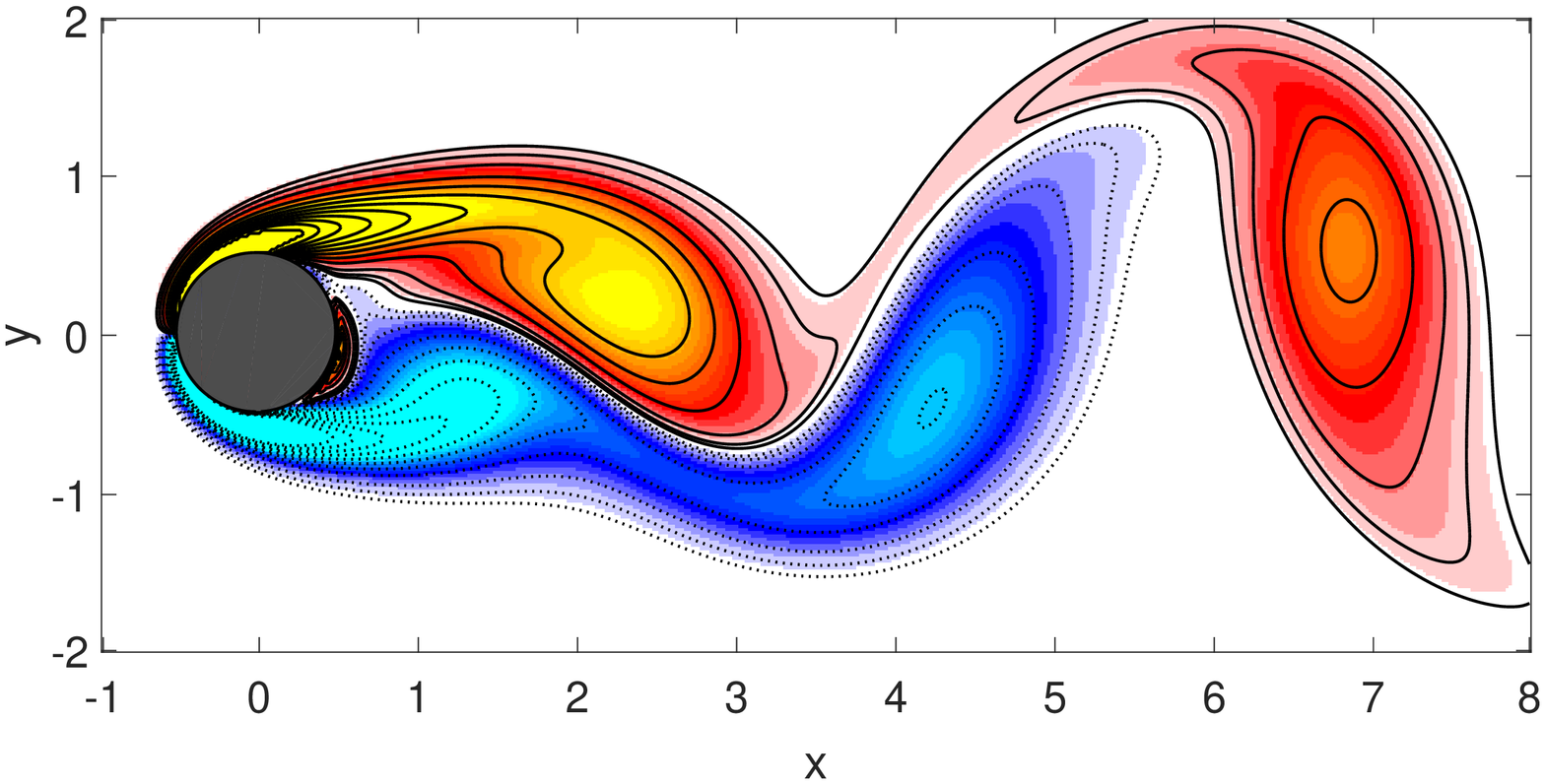}
  \caption{Snapshot of flow over circular cylinder at Re=100.}
  \label{fig:cylsnap}
\end{subfigure}%
\begin{subfigure}{.5\textwidth}
  \centering
  \includegraphics[width=\linewidth]{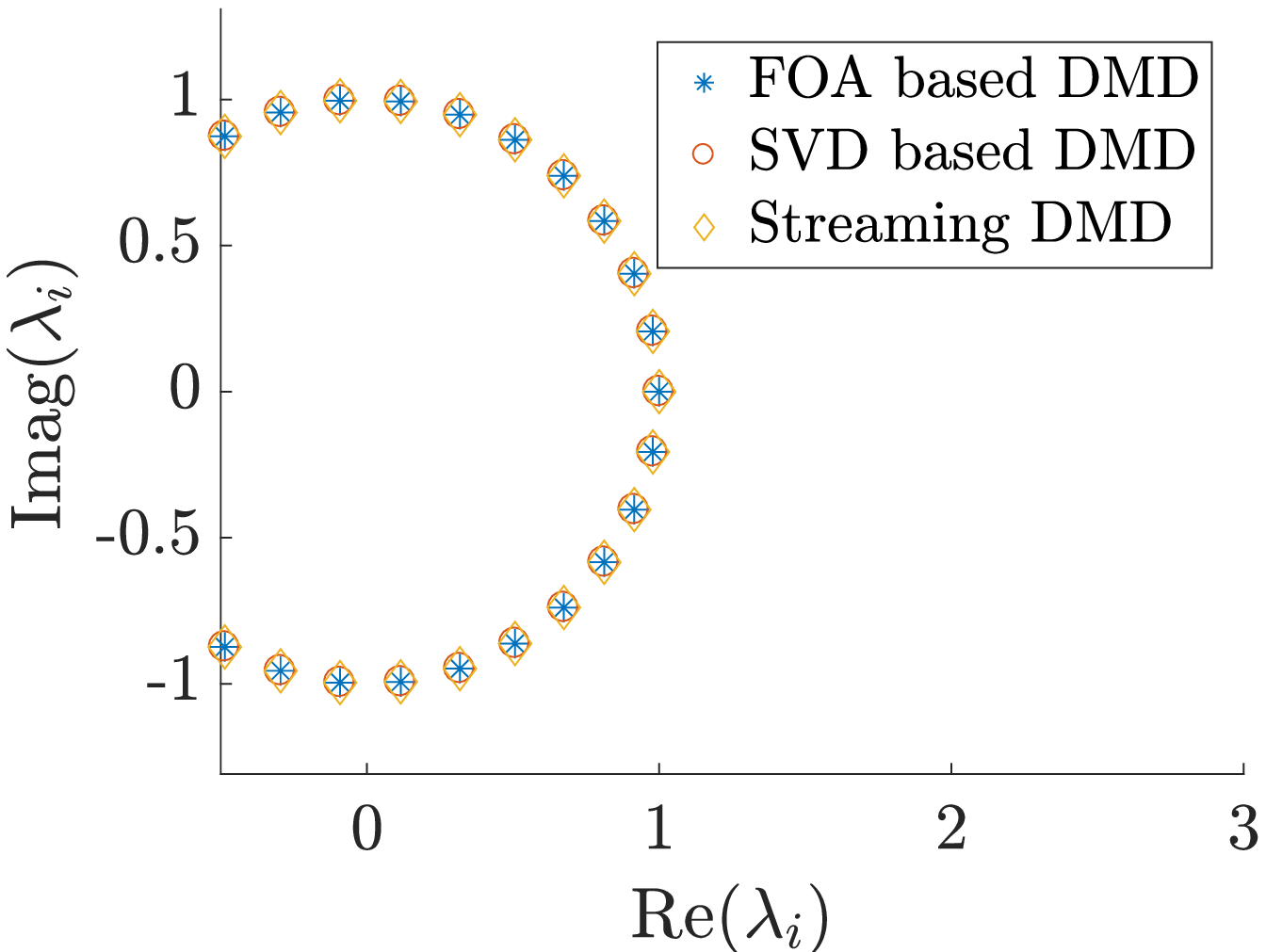}
  \caption{Comparison of DMD eigenvalues obtained from FOA based DMD, SVD based DMD \citep{schmid2010dynamic} and streaming DMD \citep{hemati2014dynamic}.}
  \label{fig:cyldmdeig}
\end{subfigure}
\caption{Cylinder test case.}
\end{figure}

We consider the flow over a circular cylinder at Reynolds number ($Re:=VD/\nu$) of 100 ($V$, $D$ and $\nu$ denote freestream velocity, diameter of cylinder and kinematic viscosity of fluid respectively). The data set for this case is obtained from \cite{kutz2016dynamic}. For details on the numerical method used to generate the cylinder snapshots refer \cite{kutz2016dynamic}. A total of 151 snapshots of vorticity field separated by time interval $\Delta t=0.2$ were used. Fig. \ref{fig:cylsnap} shows a snapshot of the vorticity field. The cylinder is placed at (0,0) and the diameter of the cylinder is 1. The flow is from left to right. It is known that as we increase $Re$, for $Re>47$ vortices are shed from the cylinder at a particular frequency dependent on $Re$. These vortices can be clearly seen from fig. \ref{fig:cylsnap}. At $Re=100$, the Strouhal number $St=fD/V$ is 0.16. We can capture this frequency by performing DMD of the snapshots of vorticity field. The condition number of the snapshots for flow over cylinder at Re=100 is $\approx 10^7$.

Fig. \ref{fig:cyldmdeig} compares DMD eigenvalues associated with 21 dominant eigenmodes from the proposed FOA based DMD method (without truncation), SVD based DMD (with rank truncation) and streaming DMD. The DMD eigenmodes from FOA based DMD are ranked based on the magnitude of $c_j$ when the first snapshot $\psi_1$ is represented in the basis of DMD eigenmodes (equation \ref{eq:exrecon}). Also, only few of the computed DMD eigenmodes have sufficiently large values of $c_j$ (not shown). These DMD eigenmodes give an appropriate basis to represent the dataset. In SVD based DMD DMD, only the first 21 singular vectors are used to evaluate the DMD eigenvalues and eigenmodes. Since, the condition number of the snapshot matrix is $\approx 10^7\ll \epsilon_m$ streaming DMD returns reliable eigenvalues. Similar to FOA based DMD, the DMD modes obtained from streaming DMD are sorted based on the magnitude of the coefficients when the first snapshot $\psi_1$ is represented as their linear combination. Figure \ref{fig:cyldmdeig} shows good agreement of the computed DMD eigenvalues between the different projected DMD methodologies considered.

\begin{figure}[]
\centering
\begin{subfigure}{.5\textwidth}
  \centering
  \includegraphics[width=\linewidth]{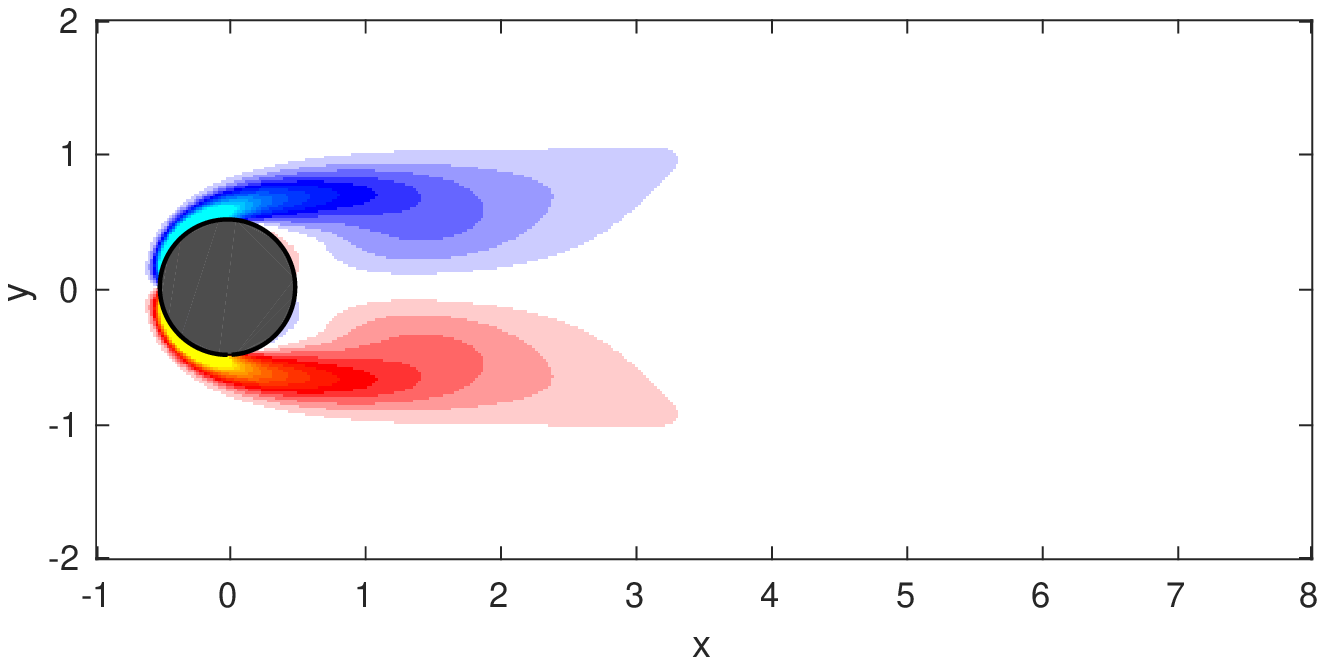}
  \caption{DMD eigenmode for $\frac{log(\lambda)}{2\pi\Delta t}=0$.}
  \label{fig:cyldmdmode1}
\end{subfigure}%
\begin{subfigure}{.5\textwidth}
  \centering
  \includegraphics[width=\linewidth]{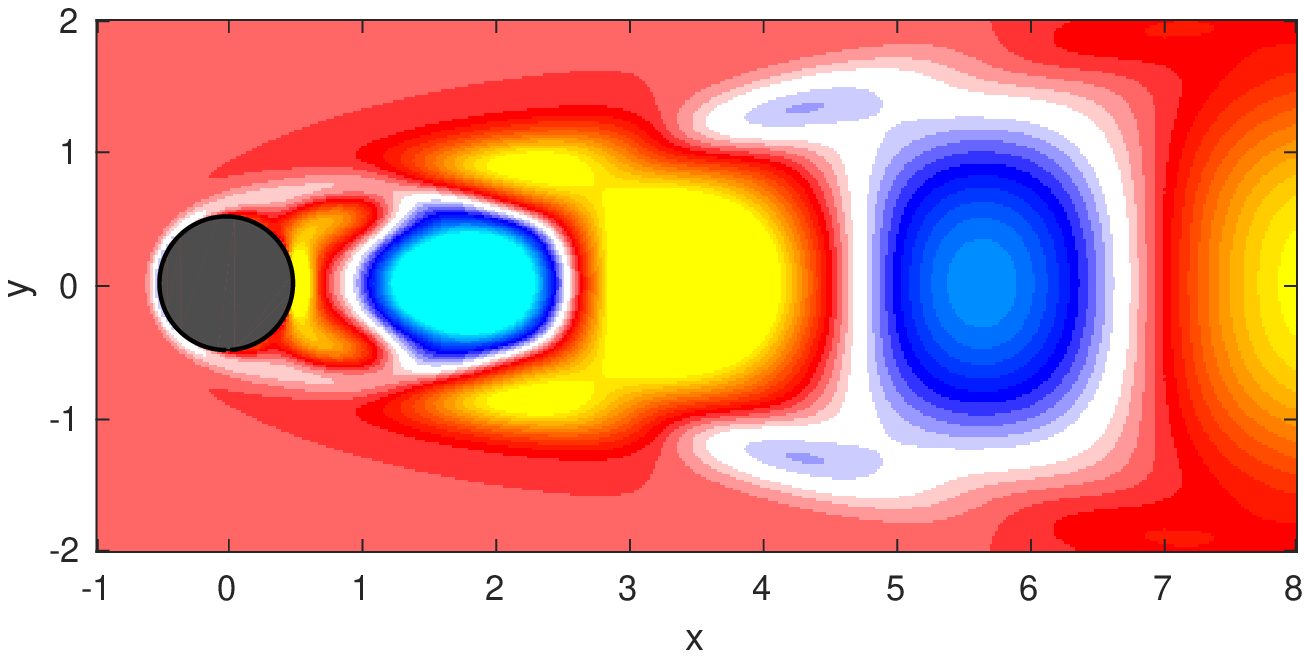}
  \caption{Real part of DMD eigenmode for $\frac{log(\lambda)}{2\pi\Delta t}=0.1654i$.}
  \label{fig:cyldmdmode2}
\end{subfigure}
\caption{DMD modes from snapshots of vorticity of flow over cylinder using FOA based DMD without rank truncation.}
\end{figure}

Figs. \ref{fig:cyldmdmode1} and \ref{fig:cyldmdmode2} show the first 2 dominant DMD eigenmodes obtained from FOA based DMD. DMD eigenmode in \ref{fig:cyldmdmode1} has DMD eigenvalue of 1 and represents the mean vorticity field. The second dominant DMD eigenmode shown in figure \ref{fig:cyldmdmode2} corresponds to the vortex shedding frequency. The frequency of oscillation of DMD mode is given by $imag(log(\lambda)/(2\pi\Delta t))$ where $\lambda$ is corresponding DMD eigenvalue. Fig. \ref{fig:cyldmdmode2} corresponds to DMD eigenvalue, $\lambda = 0.9875-0.2063i$ whose associated frequency is $0.1654$, which is exactly the $\textit{St}$ at $Re=100$.

The snapshots matrix is full rank (measured using numerical rank). So, rank truncation is not required for this case. Table \ref{tab:errorIndCyl} shows the quality of the five most dominant DMD eigenvectors and eigenvalues. Since, the condition number of the snapshots is very small when compared with $\frac{1}{\epsilon_m}$, the computed error indicator $|h_{N,N-1}e_{N-1}^Hz|$ should be close to the actual error $\|AV_1^{N-1}z-\lambda V_1^{N-1}z\|_2$.

\begin{table}
\centering
\begin{tabular}{|c|l|}
\hline
$\frac{log(\lambda)}{2\pi\Delta t}$ & $Error_{DMD}$ \\ \hline
-0.0000+(0.0000)i & 3.24e-08 \\ 
0.0000+(0.1654)i & 7.60e-08 \\ 
0.0000+(-0.1654)i & 7.60e-08 \\ 
-0.0000+(0.3308)i & 2.18e-07 \\ 
-0.0000+(-0.3308)i & 2.18e-07 \\ 
\hline
\end{tabular}
\caption{Predicted error in DMD eigenvector and eigenvalues for the 5 most dominant eigenvalues.}
\label{tab:errorIndCyl}
\end{table}

\subsection{Jets in cross flow}
Next, we consider a large dataset obtained from Direct Numerical Simulation (DNS) of turbulent jets in cross flow. DMD on this dataset has been previously performed by \cite{iyer2016numerical} using the algorithm of \cite{schmid2010dynamic} using snapshot matrix as the basis vectors. Here, we use FOA based DMD to obtain the DMD modes and eigenvalues and compare to previously obtained results. For more information on the problem, simulation, timestep and grid, the reader is referred to \cite{iyer2016numerical}. Here, we consider jet velocity to cross-flow velocity ratios (R) of 2 and 4. The size of each snapshot is 240 million. A total of 80 snapshots were used for R=4, and 250 snapshots were used for R=2 case.

\begin{figure}[]
\centering
\begin{subfigure}{.4\textwidth}
 \centering
 \includegraphics[width=\linewidth]{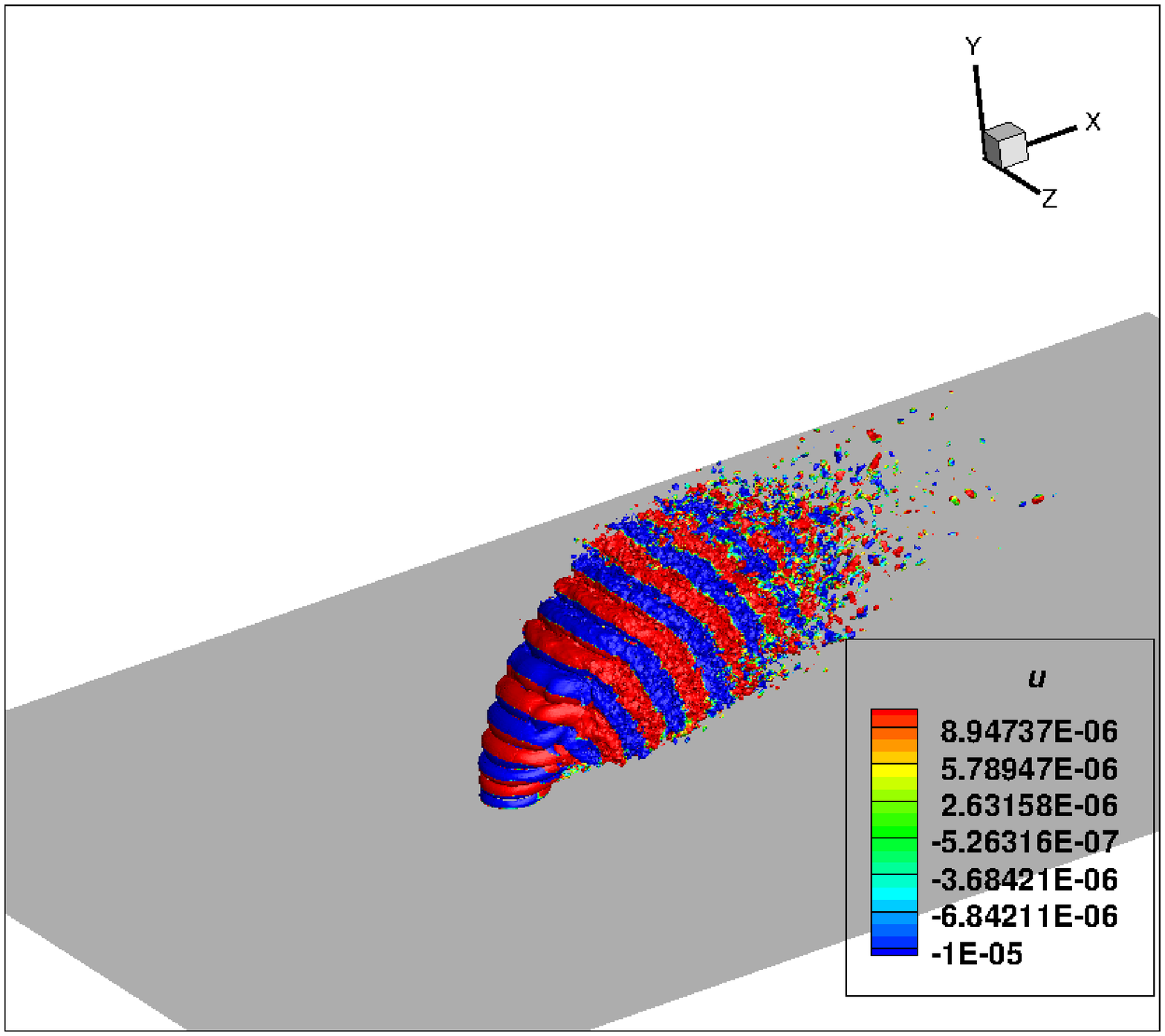}
 \caption{R=2, St=0.63}
\end{subfigure}
\begin{subfigure}{.4\textwidth}
 \centering
 \includegraphics[width=\linewidth]{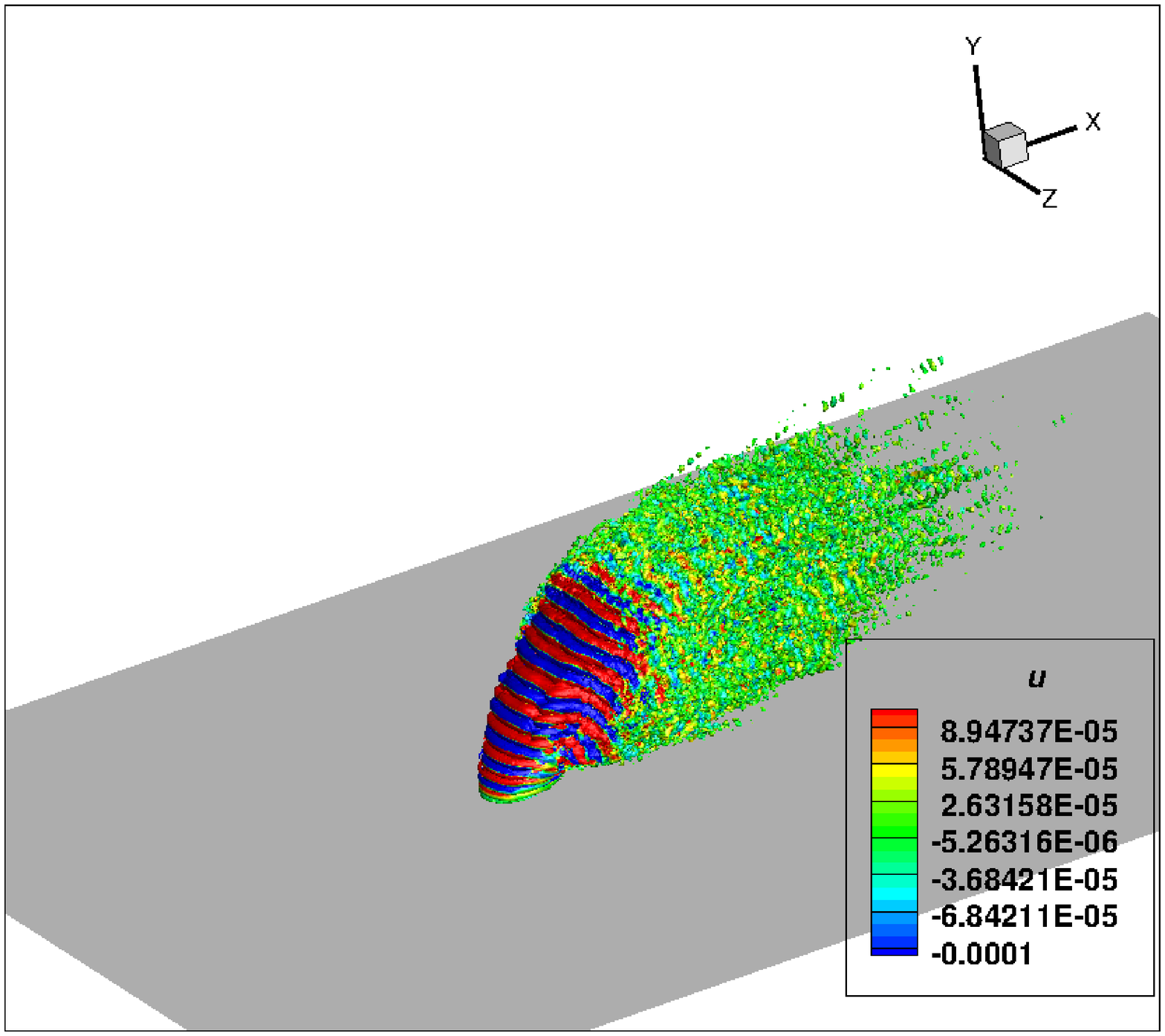}
 \caption{R=2, St=1.21}
\end{subfigure}
\begin{subfigure}{.4\textwidth}
 \centering
 \includegraphics[width=\linewidth]{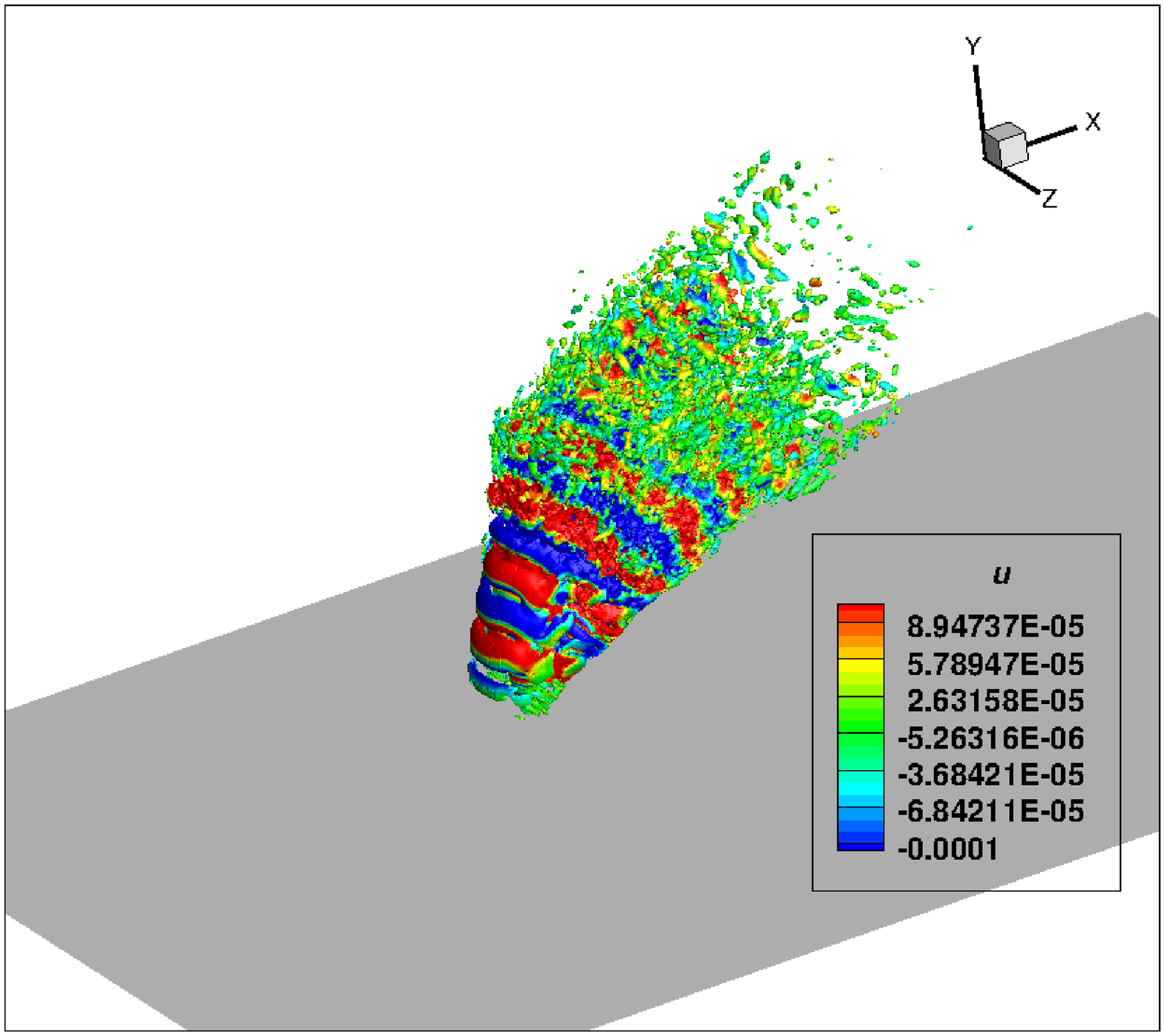}
 \caption{R=4, St=0.38}
\end{subfigure}
\begin{subfigure}{.4\textwidth}
 \centering
 \includegraphics[width=\linewidth]{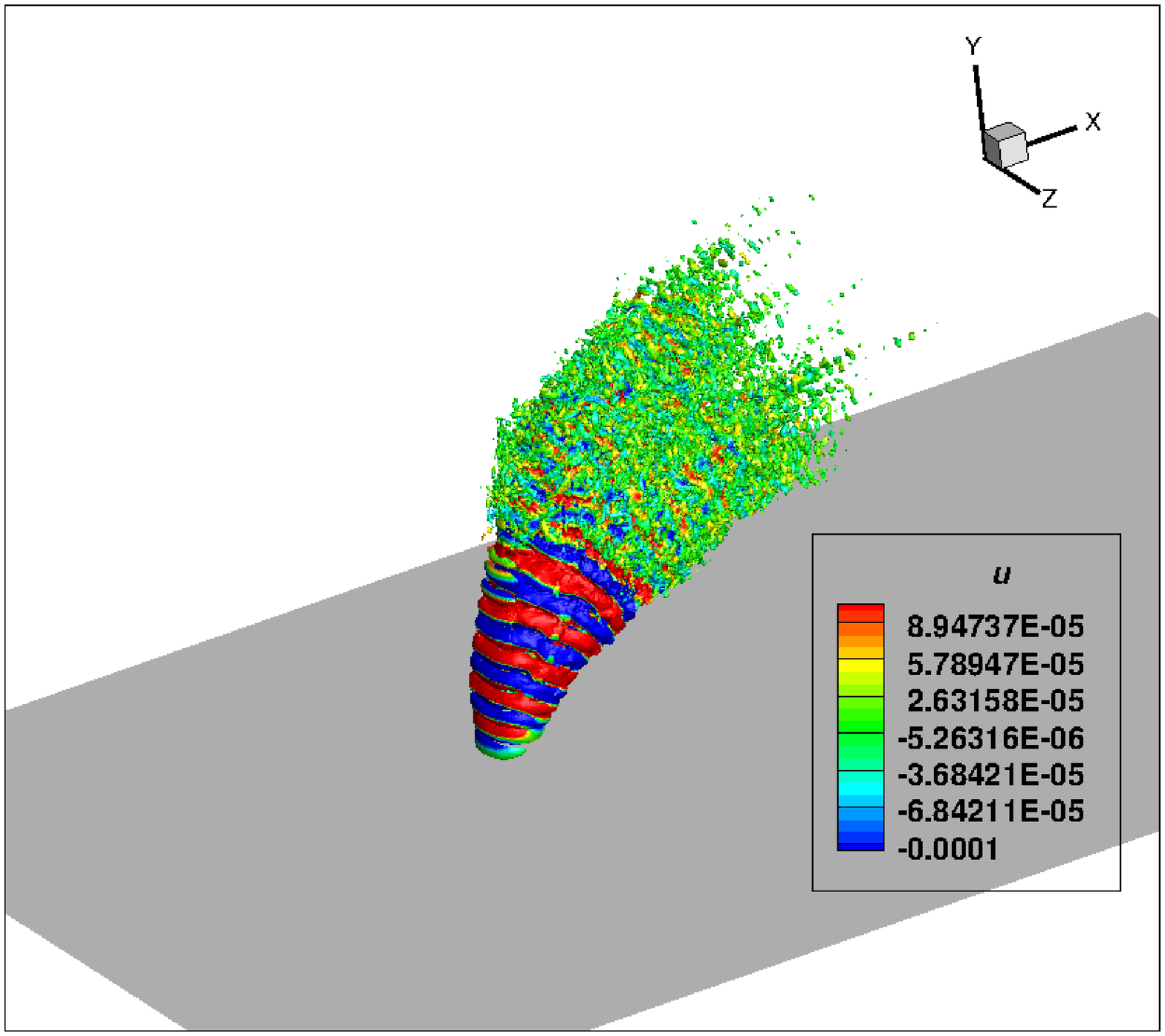}
 \caption{R=4, St=0.76}
\end{subfigure}
\caption{Real part of DMD eigenmode associated with shear layer.}
\label{fig:allmodesjicf}
\end{figure}

The parallel implementation discussed in section \ref{sec:parallelscaling} was used. The data was split row-wise among different processors. A total of 512 processors were used to process the data. The condition number of snapshot matrix is $\approx 10^2$ which is very small when compared with $1/\epsilon_m$ for double precision datatype. So, FOA based DMD without rank truncation is used to compute the DMD eigenvalues and eigenvectors. Also, in \cite{iyer2016numerical}, DMD eigenvalues and eigenvectors were obtained through the projected companion matrix. The normal equations for the last column of the companion matrix were solved using rank truncated SVD. Table \ref{tab:errorfreqjicf} shows the good agreement of the Strouhal number (non-dimensionalized with peak jet velocity and diameter of orifice) associated with shear layer modes obtained from FOA based DMD with \cite{iyer2016numerical}.  Fig. \ref{fig:allmodesjicf} shows the Q-criterion of shear layer modes at R=2 and R=4 and they agree with those obtained by \cite{iyer2016numerical}.  Also, the computed error indicator $|h_{N,N-1}e^H_{N-1}z|$ presented in Table \ref{tab:errorfreqjicf} shows the better quality of eigenvectors for R=2 case than R=4. This is because of a larger set of snapshots used for R=2 than R=4. These error indicators are very useful as it allows us to quantitatively compare the quality of DMD modes and eigenvalues from different datasets and also informs the user when to stop adding new snapshots. The finite precision error analysis of error indicators elucidates that their reliability primarily depends on the condition number of the snapshot matrix. In this case, since the condition number is reasonable, the predicted error should be very close to the actual error. The time taken to generate the Arnoldi vectors and projected matrix from the input snapshot matrix is approximately 234 seconds for 250 snapshots and 25 seconds for 80 snapshots on Stampede2 Skylake cluster.

\begin{table}
\centering
\begin{tabular}{|c|c|c|c|}
\hline
R & \citep{iyer2016numerical} & FOA based DMD & $Error_{DMD}$ \\ \hline
2 & 0.6255 & 0.6263 & 0.0194 \\
2 & 1.2077 & 1.2071 & 0.0323 \\
4 & 0.3804 & 0.3804 & 0.0545 \\
4 & 0.7624 & 0.7621 & 0.0518 \\
\hline
\end{tabular}
\caption{Comparison of Strouhal number associated with shear layer modes obtained from FOA based DMD and result of \cite{iyer2016numerical} for different jet velocity to cross flow velocity ratios (R) and the predicted error associated with DMD eigenvalues and eigenvectors.}
\label{tab:errorfreqjicf}
\end{table}

\section{Conclusion}
In this paper, we develop a modified version of the standard FOA method, which forms the kernel of the proposed FOA based DMD algorithm (Algorithm \ref{algo:foabaseddmd}). The streaming form of the proposed methodology is shown in Algorithm \ref{algo:foabaseddmdstr}. The cost and memory consumption of FOA based DMD is smaller than that of SVD based DMD \citep{schmid2010dynamic} and streaming DMD \citep{hemati2014dynamic}. From finite precision error analysis of the FOA based DMD, the accuracy of computed projection of the linear mapping $A$ is shown to be $O(\kappa_2(X_1^{N-1})\epsilon_m)$ which is same as that for SVD based methods. The finite precision error of streaming DMD is shown to be $O\left(\left(\kappa_2(X_1^{N-1})\right)^2\epsilon_m\right)$ at most. So, DMD methods which are theoretically equivalent can have different finite precision error. These error estimates are validated by considering snapshot matrices of logarithmically increasing condition number. For snapshots with extremely large condition numbers of $O(\frac{1}{\epsilon_m})$, rank truncation may be used within the FOA based DMD algorithm. The increase in accuracy of computed projection onto a subspace of Krylov subspace in rank truncated FOA based DMD method is explained using finite precision error analysis. Error indicators for the computed DMD eigenvectors and eigenvalues are derived for FOA based DMD with and without rank truncation. These error indicators can be used to devise stopping criterion for DMD. Exact reconstruction property and parallel implementation aspects of the proposed method are discussed. The method is easily parallelizable. Scaling of the algorithm is shown upto 16384 processors. The proposed algorithm is tested on three cases of increasing levels of dimensionality and condition number of snapshot matrix. The proposed error indicators are validated and bounds of accuracy of computed projection is assessed using linearized channel flow simulation snapshots at $Re=10000$. The DMD eigenvectors and eigenvalues are extracted from cylinder simulation snapshots at $Re=100$ using FOA based DMD and compared with those obtained by SVD based and streaming DMD methods. The capability of the method to perform DMD on very large scale datasets is shown by performing DMD of snapshots obtained from DNS of turbulent jet in cross flow at two different jet velocity to cross flow velocity ratios.

FOA based DMD algorithm is well-suited for DMD of large datasets on parallel computing platforms as it is a streaming method. Finite precision error analysis is important for DMD algorithms as they rely on snapshot vectors and not on the linear mapping. DMD algorithms which are equivalent in theory can have different finite precision error. The error indicators for modes computed from FOA based DMD provides a quantitative means to compare the accuracy of modes computed from different datasets and decide when to stop acquiring new snapshots. Since FOA based DMD relies on the Arnoldi method with full orthogonalization, the method can utilize the advancements made in Arnoldi-based methods for large eigenvalue problems \citep{saad2011numerical} in the context of DMD.

\section*{Acknowledgement}
The authors would like to acknowledge support by AFOSR under grant FA9550-15-1-0261. The scaling study was made possible through computing resources provided by Minnesota Supercomputing Institute (MSI) and Texas Advanced Computing Center (TACC). The authors would like to thank Dr. Prahladh Iyer for providing the jets in cross flow dataset. 

\section*{Appendix}
\appendix
\section{Computational cost and memory requirement} \label{app:candm}

\subsection{FOA based DMD} \label{app:foabaseddmdcandm}

First, we consider the cost of FOA based DMD without rank truncation to obtain the projected matrices for the batch processed algorithm shown in Algorithm \ref{algo:foabaseddmd}. The streaming algorithm has the same floating point operation count as the batch processed one.
\begin{equation} \label{eqn:costoffoabaseddmd}
\begin{split}
&\text{Cost of FOA based DMD without rank truncation} \\
&\approx \underbrace{3M}_{\text{step 1}}+\sum_{j=1}^{N-1}\left[ \underbrace{2(j-1)}_{\text{step 5}} + \underbrace{\sum_{i=2}^j2(j-i+1)}_{\text{step 6-8}} + \underbrace{2Mj + M}_{\text{step 9}} + \underbrace{8Mj + j}_{\text{step 10-14}} +\underbrace{3M}_{\text{step 15}} + \underbrace{2(j+1)}_{\text{step 16-18}} \right], \\
&\approx 5MN(N-1)+(4M+2)(N-1)+\frac{N(N-1)(2N-1)}{6}+3N(N-1)+3M, \\
&\approx 5MN^2+\frac{N^3}{3} \text{ (neglecting lower order terms)}.
\end{split}
\end{equation}
Note that in the last step of the above equation we have only considered terms which are cubic in the dimension of the problem. The cost of obtaining the projected matrix using FOA based DMD with rank truncation would include additional costs of SVD and matrix multiplications, i.e., 
\begin{equation} \label{eqn:costoffoabaseddmdwithranktruncation}
\begin{split}
&\text{Cost of FOA based DMD with rank truncation} \\
&=\text{Cost of FOA based DMD without rank truncation} + \text{Cost of SVD and matrix multiplication} \\
&\approx \underbrace{5MN^2+\frac{N^3}{3}}_{\text{FOA based DMD}}+\underbrace{12N^3}_{\text{step 22 \citep{golub2012matrix}}}+\underbrace{2r(N-1)^2+2r^2(N-1)}_{\text{step 23}}\\
&\approx 5MN^2+\frac{37}{3}N^3+2rN^2+2r^2N \text{ (neglecting lower order terms)},
\end{split}
\end{equation}
where $r$ is the truncated rank. The cost of optional 'scaling basis vectors' option of Algorithm \ref{algo:foabaseddmd} is small as it is not cubic in the matrix dimensions.

The memory requirement (number of floating point numbers to be stored) of streaming FOA based DMD without rank truncation involves storing the Arnoldi vectors $V_1^N$, one upper Hessenberg matrix with an additional row $\bar{H}_N$ and upper triangular matrix $\beta_N$. i.e.,
\begin{equation}
\begin{split}
&\text{Memory requirement of streaming FOA based DMD without rank truncation} \\
&= \underbrace{MN}_{V_1^N} + \underbrace{\frac{(N-1)(N+2)}{2}}_{\bar{H}_N} + \underbrace{\frac{N(N+1)}{2}}_{\beta_N} \\
&\approx MN + N^2\text{ (including only quadratic terms)}
\end{split}
\end{equation}
Note that in the implementation, new snapshots which are to be processed can be stored as columns of $V_1^N$ which are then overwritten by the computed orthonormal Arnoldi vectors. With rank truncation, additional memory requirement involves the left singular vectors, the singular values and the new projected full matrix.
\begin{equation}
\begin{split}
&\text{Memory requirement of streaming FOA based DMD with rank truncation} \\
&\approx MN+N^2+\underbrace{(N-1)^2}_{\text{left singular vectors}}+\underbrace{(N-1)}_{\text{singular values}}+\underbrace{r^2}_{\text{projected matrix}}\\
&\approx MN+2N^2+r^2\text{ (including only quadratic terms)}
\end{split}
\end{equation}

\subsection{SVD based DMD \citep{schmid2010dynamic}} \label{app:svdbaseddmdcandm}

SVD based methods involve computation of SVD and matrix multiplications to compute the projected matrix.
\begin{equation}
\begin{split}
&\text{Cost of SVD based DMD with rank truncation} \\
&=\text{Cost of SVD} + \text{Cost of matrix multiplication},\\
&\approx \underbrace{6M(N-1)^2+20(N-1)^3}_{\text{SVD \citep{golub2012matrix}}}+\underbrace{2Mr^2+2r^3+r^2+r}_{\text{matrix multiplication}},\\
&\approx 6MN^2+20N^3+2Mr^2+2r^3 \text{ (neglecting lower order terms)}.
\end{split}
\end{equation}
Here, $r$ is the truncated rank. The cost associated without rank truncation which is obtained by setting $r=N-1$ is $\approx 8MN^2+22N^3$.

SVD based DMD requires all the snapshots $X_1^N$, singular vectors $U$ and $W$, $N-1$ singular values and the full projected matrix.
\begin{equation}
\begin{split}
&\text{Memory requirement for SVD based DMD with rank truncation} \\
&= \underbrace{MN}_{X_1^N} + \underbrace{M(N-1)}_{U} + \underbrace{(N-1)^2}_{W} + \underbrace{N-1}_{\text{singular values}}+\underbrace{r^2}_{\text{projected matrix}}\\
&\approx 2MN+N^2+r^2\text{ (including only quadratic terms)}
\end{split}
\end{equation}
If rank truncation is not involved, the number of floating point numbers that needs to be stored is then $\approx 2MN^2+2N^2$.
\subsection{Streaming DMD \citep{hemati2014dynamic}} \label{app:streamingdmdcandm}

The method of \cite{hemati2014dynamic} without compression computes the projected matrix from snapshot pairs as 
\begin{equation}
\begin{split}
AX&=Y,\quad X=Q_X\tilde{X},\quad Y=Q_Y\tilde{Y}, \\
Q_X^HAQ_X&=Q_X^HQ_Y\tilde{Y}\tilde{X}^H\left(\tilde{X}\tilde{X}^H\right)^{-1}.
\end{split}
\end{equation}

The floating point operation count to compute the projection of $A$, $Q_X^HAQ_X$ comprises of two QR factorizations (with reorthogonalization), incrementally forming the matrices $\tilde{Y}\tilde{X}^H$ and $\tilde{X}\tilde{X}^H$, solving $\tilde{Y}\tilde{X}^H\left(\tilde{X}\tilde{X}^H\right)^{-1}$ and a few matrix-matrix multiplications. In particular,
\begin{equation}
\begin{split}
&\text{Cost of streaming DMD without compression} \\
&\approx\underbrace{8M(N-1)^2}_{\text{2 QR factorizations (with reorthogonalization)}} + \underbrace{\frac{2}{3}N(N-1)(2N-1)}_{\text{incrementally forming $\tilde{Y}\tilde{X}^H$ and $\tilde{X}\tilde{X}^H$}} + \underbrace{2(N-1)^3}_{\text{solving $\tilde{Y}\tilde{X}^H\left(\tilde{X}\tilde{X}^H\right)^{-1}$}} + \underbrace{2M(N-1)^2 + 2(N-1)^3}_{\text{matrix multiplications}}, \\
&\approx 10MN^2+\frac{16}{3}N^3 \text{ (neglecting lower order terms)}.
\end{split}
\end{equation}

The number of floating point numbers to be stored consists of two orthogonal matrices $Q_X$ and $Q_Y$, two upper triangular matrices $\tilde{X}$ and $\tilde{Y}$, two outer product matrices $\tilde{Y}\tilde{X}^H$ and $\tilde{X}\tilde{X}^H$ and the projected matrix. i.e.,
\begin{equation}
\begin{split}
&\text{Memory required for streaming DMD without compression}\\
&=\underbrace{2M(N-1)}_{\text{2 orthogonal matrices}}+\underbrace{N(N-1)}_{\text{2 upper triangular matrices}}+\underbrace{2(N-1)^2}_{\text{2 outer product matrices}}+\underbrace{(N-1)^2}_{\text{projected matrix}}\\
&\approx 2MN+4N^2\text{ (including only quadratic terms)}
\end{split}
\end{equation}
The authors do note that a special implementation of streaming DMD can be created for sequence of snapshots which has lower computational cost and memory requirement, but the method as is, consumes the above deduced cost and memory consumption.

It is important to note that the method of \cite{hemati2014dynamic} with compression is not the same as SVD based DMD and FOA based DMD with rank truncation. The compressed streaming DMD of \cite{hemati2014dynamic} makes an additional approximation that the rank truncated snapshot pairs are related by the same linear mapping $A$ that relates the snapshot pairs. i.e.,
\begin{equation}
\begin{split}
AX_r=Y_r+U_Y\Sigma_YW_Y^H\left(W_{Xr}W_{Xr}^H-W_{Yr}W_{Yr}^H\right)
\end{split}
\end{equation}
where $X=U_X\Sigma_XW_X^H$, $Y=U_Y\Sigma_YW_Y^H$, $U_{Xr}:=U_X(:,\textit{1:r})$, $U_{Yr}:=U_Y(:,\textit{1:r})$, $\Sigma_{Xr}:=\Sigma_X(\textit{1:r},\textit{1:r})$, $\Sigma_{Yr}:=\Sigma_Y(\textit{1:r},\textit{1:r})$, $W_{Xr}:=W_X(:,\textit{1:r})$, $W_{Yr}:=W_Y(:,\textit{1:r})$, $X_r:=U_{Xr}\Sigma_rW_{Xr}^H$ and $Y_r:=U_{Xr}\Sigma_rW_{Xr}^H$. In the compressed version of streaming DMD \citep{hemati2014dynamic}, the second term in the right hand side of the above equation is neglected and then the Galerkin projection is performed to obtain approximate eigenvectors of $A$ in the range of $U_r$. In the case of rank truncated SVD based DMD and FOA based DMD, the second term in right hand side is not neglected and is followed by Galerkin projection to obtain approximate eigenvectors in the range of $U_r$.

\section{Backward error of FOA based DMD} \label{app:foabaseddmdberr}

Rewriting Equation \ref{eqn:decompfoabaseddmd}, 
\begin{equation} \label{eqn:redecompfoabaseddmd}
A\hat{V}_1^{N-1}\hat{\beta}_{N-1}-\hat{V}_1^{N}\hat{\bar{H}}_N\hat{\beta}_{N-1} = A\left(\hat{V}_1^{N-1}\hat{\beta}_{N-1}-X_1^{N-1}\right) + \left(X_2^N-\hat{V}_1^{N}\hat{\beta}_{1:N,2:N}\right)+\hat{V}_1^N\left(\hat{\beta}_{1:N,2:N}-\hat{\bar{H}}_N\hat{\beta}_{N-1}\right)
\end{equation}
From Equation \ref{eqn:betacolumnupdate}, we see that $\beta_{1:N,2:N}$ is a matrix product of $\bar{H}_N$ and $\beta_{N-1}$. Using finite precision arithmetic result of matrix multiplication \citep{higham2002accuracy},
\begin{equation} \label{eqn:matmulhbeta}
\|\hat{\beta}_{1:N,2:N}-\hat{\bar{H}}_N\hat{\beta}_{N-1}\|_F \le \gamma_{N-1}\|\hat{\bar{H}}_N\|_F\|\hat{\beta}_{N-1}\|_F,
\end{equation}
where $\gamma_k:=\frac{k\epsilon_m}{1-k\epsilon_m}$. Next, we estimate $X_1^N-\hat{V}_1^N\hat{\beta}_N$ using finite precision inner product estimate for the first column and the finite precision arithmetic Arnoldi relation for the remaining columns.

For the first column $\psi_1-\hat{v}_{1}\hat{\beta}_{1,1}$, 
\begin{equation}
\begin{split}
&\hat{\beta}_{1,1}=fl(\|\psi\|_2)=\|\psi\|_2\left(1+\theta_{M+1}\right), \\
&\hat{v}_1=fl\left(\frac{\psi_1}{\hat{\beta}_{1,1}}\right)=\frac{\psi_1}{\|\psi_1\|_2}\frac{\left(1+\theta_1\right)}{\left(1+\theta_{M+1}\right)}=\frac{\psi_1}{\|\psi_1\|_2}\left(1+\theta_{2M+3}\right) \left(\because \frac{1+\theta_k}{1+\theta_j}=1+\theta_{k+2j}\text{ for }j>k\right),\\
&\psi_1-\hat{v}_1\hat{\beta}_{1,1}=-\psi_1\theta_{3M+4},
\end{split}
\end{equation}
where $\theta_k$ is a real number such that $|\theta_k|\le\gamma_k$.

For the remaining columns of $X_1^N-\hat{V}_1^N\hat{\beta}_N$, we first recognize that FOA based DMD is QR factorization of the matrix $\left[v_1,L_1,\dots,L_{N-1}\right]$ in exact arithmetic, where $L_j:=\frac{1}{\beta_{j,j}}\left(\psi_{j+1}-v_1\sum_{i=1}^{j-1}h_{1,i}\beta_{i,j}-\sum_{k=2}^jv_k\sum_{i=k-1}^{j-1}h_{k,i}\beta_{i,j}\right)$. From Equation \ref{eqn:qrfactbstab}, we have in finite precision arithmetic,
\begin{equation}\label{eqn:berrear}
\begin{split}
&\left[\hat{v}_1,\hat{L}_1,\dots,\hat{L}_{N-1}\right] = \hat{V}_1^N\hat{R}_N+\left[0,E^{AR}\right],\\
&\|E^{AR}\|_F\le c_1\epsilon_m\|\hat{L}_1,\dots,\hat{L}_{N-1}\|_F,
\end{split}
\end{equation}
where, $\hat{L}_j:=fl\left(\frac{1}{{\hat{\beta}}_{j,j}}\left(\psi_{j+1}-\hat{v}_1fl\left(\sum_{i=1}^{j-1}\hat{h}_{1,i}\hat{\beta}_{i,j}\right) -\sum_{k=2}^j\hat{v}_kfl\left(\sum_{i=k-1}^{j-1}\hat{h}_{k,i}\hat{\beta}_{i,j}\right)\right)\right)$ is the computed counterpart of $L_j$, $\hat{R}_N:=[e_1,\hat{\bar{H}}_N]$ where $e_1\in \mathbb{R}^N$ is the first canonical basis vector and $j^{th}$ column of $E^{AR}$ is the error in the $j^{th}$ step of Arnoldi. i.e.,
\begin{equation}
fl\left(\frac{1}{{\hat{\beta}}_{j,j}}\left(\psi_{j+1}-\hat{v}_1fl\left(\sum_{i=1}^{j-1}\hat{h}_{1,i}\hat{\beta}_{i,j}\right) -\sum_{k=2}^j\hat{v}_kfl\left(\sum_{i=k-1}^{j-1}\hat{h}_{k,i}\hat{\beta}_{i,j}\right)\right)\right) = \sum_{i=1}^{j+1}\hat{h}_{i,j}\hat{v}_i+e^{AR}_{j},
\end{equation}
where $e^{AR}_j$ is the $j^{th}$ column of $E^{AR}$. Separating the finite precision error in the evaluation of left hand side as 
\begin{equation}
\begin{split}
e^{LHS}_j:=\frac{1}{{\hat{\beta}}_{j,j}}\left(\psi_{j+1}-\hat{v}_1fl\left(\sum_{i=1}^{j-1}\hat{h}_{1,i}\hat{\beta}_{i,j}\right) -\sum_{k=2}^j\hat{v}_kfl\left(\sum_{i=k-1}^{j-1}\hat{h}_{k,i}\hat{\beta}_{i,j}\right)\right)\\
-fl\left(\frac{1}{{\hat{\beta}}_{j,j}}\left(\psi_{j+1}-\hat{v}_1fl\left(\sum_{i=1}^{j-1}\hat{h}_{1,i}\hat{\beta}_{i,j}\right) -\sum_{k=2}^j\hat{v}_kfl\left(\sum_{i=k-1}^{j-1}\hat{h}_{k,i}\hat{\beta}_{i,j}\right)\right)\right),
\end{split}
\end{equation}
we have 
\begin{equation}
\frac{1}{{\hat{\beta}}_{j,j}}\left(\psi_{j+1}-\hat{v}_1fl\left(\sum_{i=1}^{j-1}\hat{h}_{1,i}\hat{\beta}_{i,j}\right) -\sum_{k=2}^j\hat{v}_kfl\left(\sum_{i=k-1}^{j-1}\hat{h}_{k,i}\hat{\beta}_{i,j}\right)\right) = \sum_{i=1}^{j+1}\hat{h}_{i,j}\hat{v}_i+e^{AR}_j+e^{LHS}_j.
\end{equation}
Rearranging, adding and subtracting $\sum_{i=1}^{j+1}\hat{\beta}_{i,j+1}\hat{v}_i$,
\begin{equation}
\begin{split}
\psi_{j+1}-\sum_{i=1}^{j+1}\hat{\beta}_{i,j+1}\hat{v}_i =& \left( -\hat{\beta}_{1,j+1}+fl\left(\sum_{i=1}^{j-1}\hat{h}_{1,i}\hat{\beta}_{i,j}\right) + \hat{h}_{1,j}\hat{\beta}_{j,j}\right)\hat{v}_1\\
&+\sum_{k=2}^j\left(-\hat{\beta}_{k,j+1}+fl\left(\sum_{i=k-1}^{j-1}\hat{h}_{k,i}\hat{\beta}_{i,j}\right)+\hat{h}_{i,j}\hat{\beta}_{j,j} \right)\hat{v}_k\\
&+\left(-\hat{\beta}_{j+1,j+1}+\hat{h}_{j+1,j}\hat{\beta}_{j,j} \right)\hat{v}_{j+1}\\
&+e^{AR}_j\hat{\beta}_{j,j}+e^{LHS}_{j}\hat{\beta}_{j,j}.
\end{split}
\end{equation}
Estimating the size of each entry in $\psi_{j+1}-\sum_{i=1}^{j+1}\hat{\beta}_{i,j+1}\hat{v}_i$ using $\hat{\beta}_{i,j+1}=fl\left(fl\left(\sum_{k=1}^{j-1}\hat{h}_{i,k}\hat{\beta}_{k,j}\right)+fl\left(\hat{h}_{i,j}\hat{\beta}_{j,j}\right)\right);i=1,\dots,j$, $\hat{\beta}_{j+1,j+1}=fl\left(\hat{h}_{j+1,j}\hat{\beta}_{j,j}\right)$ and triangle inequality,
\begin{equation}
\begin{split}
|\psi_{j+1}-\sum_{i=1}^{j+1}\hat{\beta}_{i,j+1}\hat{v}_i|\le& \left(|fl\left(\sum_{i=1}^{j-1}\hat{h}_{1,i}\hat{\beta}_{i,j}\right)|+|\hat{h}_{1,j}||\hat{\beta}_{j,j}|\right)\gamma_2|\hat{v}_1|\\
&+\sum_{k=2}^j\left(|fl\left(\sum_{i=k-1}^{j-1}\hat{h}_{k,i}\hat{\beta}_{i,j}\right)|+|\hat{h}_{i,j}||\hat{\beta}_{j,j}|\right)\gamma_2|\hat{v}_k|\\
&+|\hat{h}_{j+1,j}||\hat{\beta}_{j,j}|\epsilon_m|\hat{v}_{j+1}|\\
&+|e^{AR}_j||\hat{\beta}_{j,j}|+|e_j^{LHS}||\hat{\beta}_{j,j}|.
\end{split}
\end{equation}
Then, using the finite precision dot product error estimates $|fl\left(\sum_{i=1}^{j-1}\hat{h}_{1,i}\hat{\beta}_{i,j}\right)-\sum_{i=1}^{j-1}\hat{h}_{1,i}\hat{\beta}_{i,j}|\le \gamma_{j-1}\sum_{i=1}^{j-1}|\hat{h}_{1,i}||\hat{\beta}_{i,j}|$ and $|fl\left(\sum_{i=k-1}^{j-1}\hat{h}_{k,i}\hat{\beta}_{i,j}\right)-\sum_{i=k-1}^{j-1}\hat{h}_{k,i}\hat{\beta}_{i,j}|\le \gamma_{j-k+1}\sum_{i=k-1}^{j-1}|\hat{h}_{k,i}||\hat{\beta}_{i,j}|;k=2,\dots,j$,
\begin{equation} \label{eqn:fpefoabaseddmdest1}
|\psi_{j+1}-\sum_{i=1}^{j+1}\hat{\beta}_{i,j+1}\hat{v}_i|\le \gamma_2\sum_{i=1}^j|\hat{h}_{1,i}||\hat{\beta}_{i,j}||\hat{v}_1| +\gamma_2\sum_{k=2}^j\sum_{i=k-1}^{j}|\hat{h}_{k,i}||\hat{\beta}_{i,j}||\hat{v}_k|+\epsilon_m|\hat{h}_{j+1,j}||\hat{\beta}_{j,j}||\hat{v}_{j+1}|+|e^{AR}_j||\hat{\beta}_{j,j}|+|e^{LHS}_j||\hat{\beta}_{j,j}| + O\left(\epsilon_m^2\right).
\end{equation}
Since, $e^{LHS}_j$ is the finite precision error in the addition and division operations in computing $L_j$, we have
\begin{equation} \label{eqn:fpefoabaseddmdestlhs}
|\hat{\beta}_{j,j}||e^{LHS}_j| \le \left(|\psi_{j+1}|+\sum_{i=1}^{j-1}|\hat{h}_{1,i}||\hat{\beta}_{i,j}||\hat{v}_1| +\sum_{k=2}^j\sum_{i=k-1}^{j-1}|\hat{h}_{k,i}||\hat{\beta}_{i,j}||\hat{v}_k|\right)\gamma_{j+2}.
\end{equation}
Using Equation \ref{eqn:fpefoabaseddmdestlhs} in \ref{eqn:fpefoabaseddmdest1},
\begin{equation}
\begin{split}
|\psi_{j+1}-\sum_{i=1}^{j+1}\hat{\beta}_{i,j+1}\hat{v}_i|\le& \gamma_2\sum_{i=1}^j|\hat{h}_{1,i}||\hat{\beta}_{i,j}||\hat{v}_1| +\gamma_2\sum_{k=2}^j\sum_{i=k-1}^{j}|\hat{h}_{k,i}||\hat{\beta}_{i,j}||\hat{v}_k|+\epsilon_m|\hat{h}_{j+1,j}||\hat{\beta}_{j,j}||\hat{v}_{j+1}|+|e^{AR}_j||\hat{\beta}_{j,j}|\\
&+\left(|\psi_{j+1}|+\sum_{i=1}^{j-1}|\hat{h}_{1,i}||\hat{\beta}_{i,j}||\hat{v}_1| +\sum_{k=2}^j\sum_{i=k-1}^{j-1}|\hat{h}_{k,i}||\hat{\beta}_{i,j}||\hat{v}_k|\right)\gamma_{j+2} + O\left(\epsilon_m^2\right);j=1,\dots,N-1.
\end{split}
\end{equation}
In matrix form, we then have,
\begin{equation} \label{eqn:entrywiseineqx2n}
|X_2^N-\hat{V}_1^N\hat{\beta}_{1:N,2:N}|\le \gamma_2|\hat{V}_1^N||\hat{\bar{H}}_N||\hat{\beta}_{N-1}|+|E^{AR}||\hat{\beta}^d_{N-1}|+\gamma_{N+1}\left(|X_2^N|+|\hat{V}_1^{N-1}||\hat{H}_{N-1}||\hat{\beta}^U_{N-1}|\right)+O(\epsilon_m^2)
\end{equation}
and
\begin{equation} \label{eqn:entrywiseineqx1nm1}
|X_1^{N-1}-\hat{V}_1^{N-1}\hat{\beta}_{N-1}|\le \left[\gamma_{3M+4}|\psi_1|,\gamma_2|\hat{V}_1^{N-1}||\hat{\bar{H}}_{N-1}||\hat{\beta}_{N-2}|+|E^{AR}_{:,1:N-2}||\hat{\beta}^d_{N-2}|+\gamma_{N+1}\left(|X_2^{N-1}|+|\hat{V}_1^{N-2}||\hat{H}_{N-2}||\hat{\beta}^U_{N-2}|\right)\right] + O(\epsilon_m^2),
\end{equation}
where $\hat{\beta}^d_{j}$ is the diagonal matrix formed by using the diagonal entries of $\hat{\beta}_j$ and $\hat{\beta}^U_{j}$ is the strictly upper triangular matrix formed by using the strictly upper triangular entries of $\hat{\beta}_j$.

From Equation \ref{eqn:berrear} and using the assumption that $c_1\epsilon_m<1$, we can bound the backward error $E^{AR}$ using the computed matrix $\hat{\bar{H}}_N$ as
\begin{equation} \label{eqn:earhbarn}
\|E^{AR}\|_F\le \frac{c_1\epsilon_m}{1-c_1\epsilon_m}\sqrt{N}\|\hat{\bar{H}}_N\|_F.
\end{equation}
Taking Frobenius norm of Equation \ref{eqn:entrywiseineqx2n} and \ref{eqn:entrywiseineqx1nm1}, and using Equation \ref{eqn:earhbarn} we get,
\begin{equation}
\begin{split}
\|X_2^N-\hat{V}_1^{N}\hat{\beta}_{1:N,2:N}\|_F\le& \gamma_2\sqrt{N}\|\hat{\bar{H}}_N\|_F\|\hat{\beta}_{N-1}\|_F+\frac{c_1\epsilon_m}{1-c_1\epsilon_m}\sqrt{N}\|\hat{\bar{H}}_N\|_F\|\hat{\beta}_{N-1}\|_F+\\
&\gamma_{N+1}\left(\|A\|_F\|X_1^{N-1}\|_F+\sqrt{N-1}\|\hat{H}_{N-1}\|_F\|\hat{\beta}_{N-1}\|_F\right)+O\left(\epsilon_m^2\right)
\end{split}
\end{equation}
and 
\begin{equation} \label{eqn:errorinfact}
\begin{split}
\|X_1^{N-1}-\hat{V}_1^{N-1}\hat{\beta}_{N-1}\|_F\le& \gamma_{3M+4}\|X_1^{N-1}\|_F+\gamma_2\sqrt{N-1}\|\hat{\bar{H}}_{N-1}\|_F\|\hat{\beta}_{N-2}\|_F+\\
&\frac{c_1\epsilon_m}{1-c_1\epsilon_m}\sqrt{N-1}\|\hat{\bar{H}}_{N-1}\|_F\|\hat{\beta}_{N-2}\|_F+\gamma_{N+1}\sqrt{N-2}\|\hat{H}_{N-2}\|_F\|\hat{\beta}_{N-2}\|_F+O\left(\epsilon_m^2\right)
\end{split}
\end{equation}

From Equation \ref{eqn:errorinfact}, we see that $\|\beta_{N-1}\|_F=\|X_1^{N-1}\|_2\sqrt{N-1}+O\left(\epsilon_m\right)$ as $\hat{V}_1^{N-1}$ is orthonormal upto machine precision and since $\|\hat{\beta}_{N-2}\|_F\le\|\hat{\beta}_{N-1}\|_F$,
\begin{equation} \label{eqn:foabaseddmddecomperr}
\begin{split}
\|\hat{\beta}_{1:N,2:N}-\hat{\bar{H}}_N\hat{\beta}_{N-1}\|_F \le& \gamma_{N-1}\sqrt{N-1}\|\hat{\bar{H}}_N\|_F\|X_1^{N-1}\|_2+O(\epsilon_m^2),\\
\|X_2^N-\hat{V}_1^{N}\hat{\beta}_{1:N,2:N}\|_F\le& \gamma_2N\|\hat{\bar{H}}_N\|_F\|X_1^{N-1}\|_2+\frac{c_1\epsilon_m}{1-c_1\epsilon_m}N\|\hat{\bar{H}}_N\|_F\|X_1^{N-1}\|_2+\\
&\gamma_{N+1}\left(\left(N-1\right)\|A\|_F\|X_1^{N-1}\|_2+\left(N-1\right)\|\hat{H}_{N-1}\|_F\|X_1^{N-1}\|_2\right)+O\left(\epsilon_m^2\right),\\
\|X_1^{N-1}-\hat{V}_1^{N-1}\hat{\beta}_{N-1}\|_F\le& \gamma_{3M+4}\|X_1^{N-1}\|_F+\gamma_2\left(N-1\right)\|\hat{\bar{H}}_{N-1}\|_F\|X_1^{N-1}\|_2+\\
&\frac{c_1\epsilon_m}{1-c_1\epsilon_m}\left(N-1\right)\|\hat{\bar{H}}_{N-1}\|_F\|X_1^{N-1}\|_2+\gamma_{N+1}\left(N-2\right)\|\hat{H}_{N-2}\|_F\|X_1^{N-1}\|_2+O\left(\epsilon_m^2\right).
\end{split}
\end{equation}
Using the above equation in Equation \ref{eqn:redecompfoabaseddmd} we get,
\begin{equation}
\|A\hat{V}_1^{N-1}\hat{\beta}_{N-1}-\hat{V}_1^N\hat{\bar{H}}_N\hat{\beta}_{N-1}\|_F\le C_1\left(\|A\|_2,\|\hat{\bar{H}}_N\|_2,M,N\right)\epsilon_m\|X_1^{N-1}\|_2+O\left(\epsilon_m^2\right),
\end{equation}
where $C_1\left(\|A\|_2,\|\hat{\bar{H}}_N\|_2,M,N\right)$ is a constant which is a function of $\|A\|_2$, $\|\hat{\bar{H}}_N\|_2$, $M$ and $N$.

\bibliographystyle{model1-num-names}
\bibliography{papers}

%

\end{document}